\documentclass[11pt]{article}
\usepackage{amsmath}
\usepackage{graphicx}
\usepackage{enumerate}
\usepackage{natbib}
\usepackage{url}
\usepackage{enumitem}
\usepackage[dvipsnames]{xcolor}

\usepackage{amsmath,amssymb,amsthm,bm,mathtools}
\usepackage{algorithm}
\usepackage{dsfont,multirow,hyperref,setspace,enumerate}
\hypersetup{colorlinks,linkcolor={blue},citecolor={blue},urlcolor={black}}
\usepackage{lscape}
\usepackage{afterpage}

\allowdisplaybreaks
\usepackage{url} 
 
\theoremstyle{definition}
\newtheorem{thm}{Theorem}[section]
\newtheorem{lem}{Lemma}
\newtheorem{prop}{Proposition}[section]

\theoremstyle{definition}
\newtheorem{assumption}{Assumption}
\newtheorem{defn}{Definition}
\newtheorem{example}{Example}[section]
\newtheorem{rmk}{Remark}[section]

\newtheorem{innercustomgeneric}{\customgenericname}
\providecommand{\customgenericname}{}
\newcommand{\newcustomtheorem}[2]{%
  \newenvironment{#1}[1]
  {%
   \renewcommand\customgenericname{#2}%
   \renewcommand\theinnercustomgeneric{##1}%
   \innercustomgeneric
  }
  {\endinnercustomgeneric}
}

\newcustomtheorem{customexample}{Example}

\usepackage{appendix}
\usepackage{wrapfig}
\mathtoolsset{showonlyrefs}

\usepackage{stmaryrd}

\def\bbR{\mathbb{R}}

\def\bbE{\mathbb{E}}

\def\mGamma{\bm{\Gamma}}

\def\me{\bm{e}}

\def\mr{\bm{r}}

\def\mw{\bm{w}}
\def\mx{\bm{x}}
\def\my{\bm{y}}
\def\mz{\bm{z}}

\def\mA{\bm{A}}
\def\mB{\bm{B}}
\def\mC{\bm{C}}

\def\mI{\bm{I}}

\def\mI{\bm{I}}
\def\mM{\bm{M}}

\def\mO{\bm{O}}
\def\mP{\bm{P}}
\def\mQ{\bm{Q}}
\def\mR{\bm{R}}

\def\mU{\bm{U}}
\def\mV{\bm{V}}
\def\mW{\bm{W}}
\def\mX{\bm{X}}
\def\mY{\bm{Y}}
\def\mZ{\bm{Z}}
\def\mPhi{\bm{\Phi}}

\def\trueB{\mathcal{B}_{\textup{true}}}

\def\trueT{\Theta_{\textup{true}}}
\def\trueM{\mM_{k,\textup{true}}}

\def\mx{\bm x}
\def\mZ{\bm Z}
\def\mA{\bm A}
\def\mB{\bm B}

\def\mC{\bm C}
\def\mI{\bm I}

\def\mX{\bm X}
\def\mY{\bm Y}
\def\mM{\bm M}
\def\mSigma{\bm \Sigma}
\def\mOmega{\bm \Omega}

\def\tA{\mathcal{A}}
\def\tB{\mathcal{B}}
\def\tC{\mathcal{C}}

\def\tE{\mathcal{E}}
\def\tF{\mathcal{F}}
\def\tG{\mathcal{G}}

\def\tL{\mathcal{L}}

\def\tN{\mathcal{N}}
\def\tO{\mathcal{O}}
\def\tP{\mathcal{P}}

\def\tS{\mathcal{S}}

\def\tX{\mathcal{X}}
\def\tY{\mathcal{Y}}

\def\entry#1{\llbracket #1 \rrbracket}
\def\MLEB{\mathcal{\hat B}_{\textup{MLE}}}

\def\entry#1{\llbracket #1 \rrbracket}

\def\bbR{\mathbb{R}}

\newcommand{\onorm}[1]{\left\lVert#1\right\rVert}

\newcommand{\vnormSize}[2]{#1\lVert#2#1\rVert_2}

\newcommand{\FnormSize}[2]{#1\lVert#2#1\rVert_F} 
\newcommand{\mnormSize}[2]{#1\lVert#2#1\rVert_\infty}

\DeclareMathOperator*{\argmax}{arg\,max}

\usepackage[english]{babel}

\newcommand*{\KeepStyleUnderBrace}[1]{
  \mathop{%
    \mathchoice
    {\underbrace{\displaystyle#1}}%
    {\underbrace{\textstyle#1}}%
    {\underbrace{\scriptstyle#1}}%
    {\underbrace{\scriptscriptstyle#1}}%
  }\limits
}

\usepackage{algpseudocode,algorithm}
\algnewcommand\algorithmicinput{\textbf{Input:}}
\algnewcommand\algorithmicoutput{\textbf{Output:}}
\algnewcommand\INPUT{\item[\algorithmicinput]}
\algnewcommand\OUTPUT{\item[\algorithmicoutput]}

\usepackage{booktabs}

\allowdisplaybreaks

\usepackage{xr}
\makeatletter
\newcommand*{\addFileDependency}[1]{
  \typeout{(#1)}
  \@addtofilelist{#1}
  \IfFileExists{#1}{}{\typeout{No file #1.}}
}
\makeatother

\begin{document}

\def\spacingset#1{\renewcommand{\baselinestretch}%
{#1}\small\normalsize} \spacingset{1}

\title{Supervised tensor decomposition with features\\ on multiple modes}
  
  \author{Jiaxin Hu, Chanwoo Lee, and Miaoyan Wang\thanks{The authors gratefully acknowledge NSF grant DMS-1915978 and funding from the Wisconsin Alumni Research Foundation. Corresponding author: Miaoyan Wang, miaoyan.wang@wisc.edu. }\hspace{.2cm}\\
    Department of Statistics, University of Wisconsin-Madison}
    
    \date{}

  \maketitle

\bigskip
\begin{abstract}
Higher-order tensors have received increased attention across science and engineering. While most tensor decomposition methods are developed for a single tensor observation, scientific studies often collect side information, in the form of node features and interactions thereof, together with the tensor data. Such data problems are common in neuroimaging, network analysis, and spatial-temporal modeling. Identifying the relationship between a high-dimensional tensor and side information is important yet challenging. Here, we develop a tensor decomposition method that incorporates multiple feature matrices as side information. Unlike unsupervised tensor decomposition, our supervised decomposition captures the effective dimension reduction of the data tensor confined to feature space of interest. An efficient alternating optimization algorithm with provable spectral initialization is further developed. Our proposal handles a broad range of data types, including continuous, count, and binary observations. We apply the method to diffusion tensor imaging data from human connectome project and multi-relational political network data. We identify the key global connectivity pattern and pinpoint the local regions that are associated with available features. Our simulation code, R-package \texttt{tensorregress}, and datasets used in the paper are available at~\url{https://CRAN.R-project.org/package=tensorregress}.

\end{abstract}

\noindent%
{\it Keywords:}  Tensor data analysis, Supervised dimension reduction, Exponential family distribution, Generalized multilinear model, Alternating optimization
\vfill

\newpage
\spacingset{1.5} 
\section{Introduction}
\label{sec:intro}
\begin{figure*}[t]
\begin{center}
\includegraphics[width=15cm]{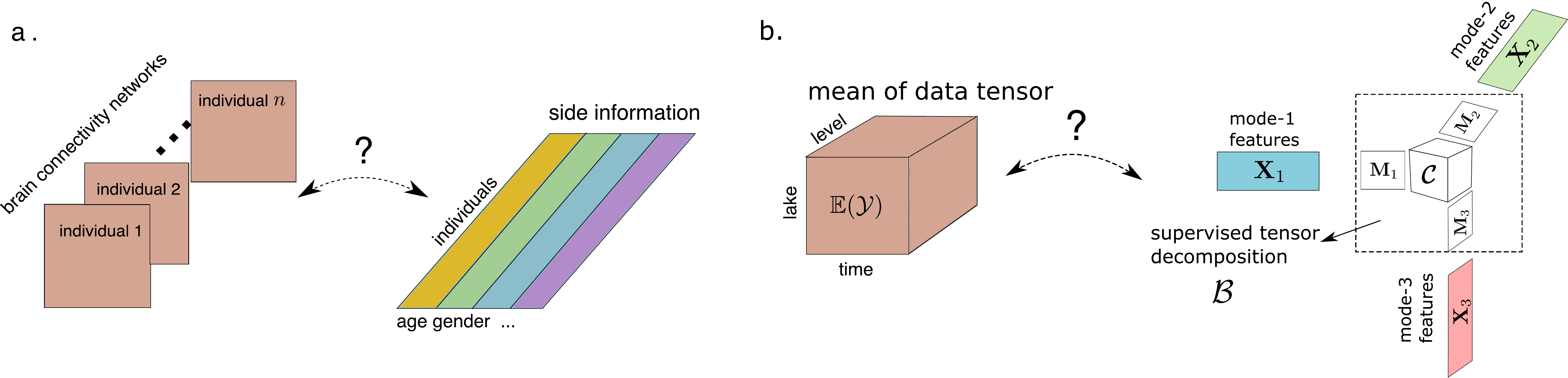}
\end{center}
\caption{Examples of supervised tensor decomposition with side information. (a) Network population model. (b) Spatio-temporal growth model.} \label{fig:intro1}
\vspace{-.2cm}
\end{figure*}

Multi-dimensional arrays, known as tensors, are often collected with side information on multiple modes in modern scientific and engineering studies. A popular example is in neuroimaging~\citep{zhou2013tensor}. The brain connectivity networks are collected from a sample of individuals, accompanied by individual characteristics such as age, gender, and diseases status (see Figure~\ref{fig:intro1}a). Another example is in network analysis~\citep{pmlr-v108-berthet20a,hoff2005bilinear}. A typical social network consists of nodes that represent people and edges that represent  friendships. Side information such as people’s demographic information and friendship types are often available. In both examples, we are interested in identifying the variation in the data tensor (e.g., brain connectivities, social community patterns) that is affected by available features. These seemingly different scenarios pose a common yet challenging problem for tensor data modeling. 

In addition to the aforementioned challenges, many tensor datasets consist of non-Gaussian measurements. Examples include the political interaction dataset \citep{nickel2011three} which measures action counts between countries under various relations, and the brain connectivity network dataset \citep{zhou2013tensor,wang2020learning} which is a collection of binary adjacency matrices. Classical tensor decomposition methods are based on minimizing the Frobenius norm of deviation, leading to suboptimal predictions for binary- or count-valued response variables. A number of supervised tensor methods have been proposed \citep{lock2018supervised,li2017parsimonious,sun2017store,hao2019sparse,raskutti2019convex} to address the tensor regression problem in various forms, such as scalar-to-tensor regression and tensor-response regression. These methods often assume Gaussian distribution for the tensor entries, or impose random designs for the feature matrices, both of which are less suitable for applications of our interest. The gap between theory and practice means a great opportunity to model paradigms and better capture the complexity in tensor data. 

We present a general model and associated method for decomposing a data tensor whose entries are from exponential family with side information. We formulate the learning task as a structured regression problem, with tensor observation serving as the response, and the multiple side information as features. Figure~\ref{fig:intro1}b illustrates our model in the special case of order-3 tensors. A low-rank structure is imposed to the conditional mean of tensor observation, where unlike classical decomposition, the tensor factors $\mX_k\mM_k\in\mathbb{R}^{d_k\times r_k}$ belong to the space spanned by features $\mX_k\in\mathbb{R}^{d_k\times p_k}$ for $k=1,2,3$. The unknown matrices $\mM_k\in\mathbb{R}^{p_k\times r_k}$ (referred to as ``dimension reduction matrices'') link the conditional mean to the feature spaces, thereby allowing the identification of variations in the tensor data attributable to the side information.

Our proposal blends the modeling power of generalized linear model (GLM) and the exploratory capability of tensor dimension reduction in order to take the best out of both worlds. We leverage GLM to allow heteroscedacity due to the mean-variance relationship in the non-Gaussian data. This flexibility is important in practice. Furthermore, our low-rank model on the (transformed) conditional mean tensor effectively mitigates the curse of high dimensionality. In classical GLM, the sample size and feature dimension are well defined; however, in the tensor data analysis, we observe only one realization of an order-$K$ tensor and up to $K$ feature matrices. Both the number of tensor entries and feature dimension grow exponentially in $K$. Dimension reduction is therefore crucial for prediction and interpretability. We establish the statistical and algorithmic convergences of our estimator, and we quantify the gain in accuracy through simulations and case studies.

Our work is closely related to but also clearly distinctive from several lines of previous work. {The first line is a class of}\label{change:firstline} \textit{unsupervised} tensor decomposition such as classical Tucker and CP decomposition~\citep{de2000multilinear, kolda2009tensor} and generalized decomposition for non-Gaussian data \citep{chi2012tensors, tarzanagh2019regularized,hong2020generalized,  li2020generalized}. {Regardless of the implementation, the unsupervised methods aim to find the best low-rank representation of a data tensor alone. In contrast, our model is a \textit{supervised} tensor learning, which aims to identify the association between a data tensor and multiple features. The low-rank factorization is determined jointly by the tensor data and feature matrices in our model.}

The second line of work studies the tensor-to-tensor regression. { This category is further divided into three scenarios, depending on whether tensor is treated as predictors~\citep{zhou2013tensor,raskutti2019convex,han2020optimal}, as responses~\citep{li2017parsimonious,sun2017store,zhang2018network,lock2018supervised,luo2018leveraging}, or both~\citep{lock2018tensor,gahrooei2020multiple}. As we show in Section~\ref{sec:connection}, our supervised tensor decomposition falls into this general category, and we provide a \emph{provable} solution in new settings that have broader practical significance. Earlier work in this vein~\citep{lock2018tensor,lock2018supervised,gahrooei2020multiple,li2020generalized} focuses on algorithm development, but not on the statistical accuracy. \cite{li2017parsimonious} introduces an envelope-based approach to identify sufficient dimension reduction~\citep{adragni2009sufficient}, but its theory is restricted to Gaussian data with one-sided feature matrix only. \cite{raskutti2019convex} establishes the statistical accuracy for convex relaxed maximum likelihood estimator (MLE) of tensor regression. However, convex relaxation for tensor optimizations suffers from computational intractability and statistical sub-optimality. Recent work has demonstrated the success of non-convex approaches in various tensor problems~\citep{sun2017store,zhang2018network,raskutti2019convex,han2020optimal}; we go step further by allowing multiple feature matrices with either fixed or random designs. In Sections~\ref{subsec:statprob}, we show that incorporating multiple feature matrices substantially improves the statistical accuracy. We provide a detailed comparison in Section~\ref{sec:connection}; see Table~\ref{table:comp_table}.}

The third line of work uses side information for various tensor learning tasks, such as for completion~\citep{song2019tensor} and for recommendation system~\citep{farias2019learning}. These methods also study tensors with side information, but they take data-mining approaches to penalize predictions that are distant from side information. One important difference is that their goal is prediction but not parameter estimation. The effects of features and their interactions are not estimated in these data-driven approaches. In contrast, our goal is interpretable prediction, and we estimate the low-rank decomposition using a model-based approach.
The model-based approaches benefits the interpretability in prediction. In this regards, our method opens up new opportunities for tensor data analysis in a wider range of applications.

The remainder of the paper is organized as follows. Section~\ref{sec:pre} introduces tensor preliminaries. Section~\ref{sec:model} presents the main model and three motivating examples for supervised tensor decomposition. We describe the likelihood estimation and alternating optimization algorithm with theoretical guarantees in Section~\ref{sec:est}. Connection with related work is provided in Section~\ref{sec:connection}. In Section~\ref{sec:simulation}, we present numerical experiments and assess the performance in comparison to alternative methods. In Section~\ref{sec:data}, we apply the method to diffusion tensor imaging data from human connectome project and multi-relational social network data. In Section~\ref{sec:proof}, we provide the proofs for the theoretical results. We conclude in Section~\ref{sec:con} with discussions about our findings and avenues of future work. 

\section{Preliminaries}\label{sec:pre}
We introduce the basic tensor properties used in the paper. We use lower-case letters (e.g.,\ $a,b,c$) for scalars and vectors, upper-case boldface letters (e.g.,\ $\mA, \mB, \mC$) for matrices, and calligraphy letters (e.g.,\ $\tA, \tB, \tC$) for tensors of order three or greater. We use $\mI$ to denote the identity matrix whose dimension may vary from line by line given the contexts. Let $\tY=\entry{y_{i_1,\ldots,i_K}}\in \mathbb{R}^{d_1\times \cdots\times d_K}$ denote an order-$K$ $(d_1,\ldots,d_K)$-dimensional tensor, where $K$ is the number of modes and also called the order. The multilinear multiplication of a tensor $\tY\in\mathbb{R}^{d_1\times \cdots\times d_K}$ by matrices $\mX_k=\entry{x_{i_k,j_k}^{(k)}}\in\mathbb{R}^{p_k\times d_k}$ is defined as 
\[
\tY \times_1\mX_1\times \cdots \times_K \mX_K=\entry{\sum_{i_1,\ldots,i_K}y_{i_1,\ldots,i_K}x_{j_1,i_1}^{(1)}\cdots x_{j_K,i_K}^{(K)}},
\]
which results in an order-$K$ $(p_1,\ldots,p_K)$-dimensional tensor. For ease of presentation, we use the shorthand $\tY\times\{\mX_1,\ldots,\mX_K\}$ to denote the tensor-by-matrix product. For any two tensors $\tY=\entry{y_{i_1,\ldots,i_K}}$, $\tY'=\entry{y'_{i_1,\ldots,i_K}}$ of identical order and dimensions, their inner product is defined as 
\[
\langle \tY,\ \tY'\rangle =\sum_{i_1,\ldots,i_K}y_{i_1,\ldots,i_K}y'_{i_1,\ldots,i_K}.
\]
The tensor Frobenius norm and maximum norm are defined as
\[
\FnormSize{}{\tY}=\langle \tY,\ \tY \rangle^{1/2},\quad \text{and}\quad \mnormSize{}{\tY}=\max_{i_1,\ldots,i_K}y_{i_1,\ldots,i_K}.
\]
When $a$ is a vector, we use $\vnormSize{}{a}=\langle a,a \rangle^{1/2}$ to denote the vector 2-norm. We use $[d]$ to denote the $d$-set $[d]=\{1,\ldots,d\}$, and use $\mathbb{O}(d,r)$ to denote the collection of all $d$-by-$r$ matrices with orthonormal columns; i.e., $\mathbb{O}(d,r)=\{\mP\in\mathbb{R}^{d\times r}\colon \mP^T\mP = \mI\}$. 

A higher-order tensor can be reshaped into a lower-order object. We use $\textup{vec}(\cdot)$ to denote the operation that reshapes the tensor into a vector, and $\text{Unfold}_k(\cdot)$ to denote the unfolding operation that reshapes the tensor along mode $k$ into a matrix of size $d_k$-by-$\prod_{i\neq k}d_i$. We use $\text{rank}(\tY)=\mr$ to denote the multilinear rank of an order-$K$ tensor $\tY$, where $\mr=(r_1,\ldots,r_K)$ is a length-$K$ vector and $r_k$ is the rank of matrix $\textup{Unfold}_k(\tY)$ for $k\in[K]$. For ease of notation, we allow the basic arithmetic operators (e.g.,\ $+, -, \geq $) and univariate functions $f\colon \mathbb{R}\to \mathbb{R}$ to be applied to tensors in an element-wise manner. For two positive sequences $\{a_n\}$ and $\{b_n\}$, we use $a_n\lesssim b_n$ or $a_n = \tO(b_n)$ to denote the fact that $a_n \leq Cb_n$ for some constant $C >0$.

\section{Motivation and model}\label{sec:model}
\subsection{General framework for tensor decomposition}\label{sec:mainmodel}
We begin with a general framework for supervised tensor decomposition and then discuss its implication in three concrete examples. Let $\tY=\entry{y_{i_1,\ldots,i_K}}\in\mathbb{R}^{d_1\times \cdots\times d_K}$ denote an order-$K$ data tensor. Suppose the side information is available on each of the $K$ modes. Let $\mX_k=\entry{x_{ij}}\in\mathbb{R}^{d_k\times p_k}$ denote the feature matrix on the mode $k\in[K]$, where $x_{ij}$ denotes the $j$-th feature value for the $i$-th tensor entity, for $(i,j)\in[d_k]\times[p_k]$. 

We propose a multilinear conditional mean model between the data tensor and feature matrices. Assume that, conditional on the features $\mX_k$, the entries of tensor $\tY$ are independent realizations from an exponential family distribution. Further, the conditional mean tensor admits the rank-$\mr$ model with $\mr=(r_1,\ldots,r_K)$,
\begin{align}\label{eq:decomp}
\mathbb{E}(\tY|\mX_1,\ldots,\mX_K) &= f\left(\tC\times\{\mX_1\mM_1, \ldots, \mX_K\mM_K\}\right),\notag\\
\text{with} \ &\ \mM_k^T\mM_k = \mI_{r_k},\ \mM_k\in\mathbb{R}^{p_k\times r_k}\quad \text{for all }k=1,\ldots,K,
\end{align}
where $\tC \in \mathbb{R}^{r_1\times \cdots \times r_K}$ is an unknown full-rank core tensor, $\mM_k \in \mathbb{R}^{p_k\times r_k}$ are unknown factor matrices for all $k\in [K]$, $f(\cdot)$ is a known link function whose form depending on the data type of $\tY$, and $\times$ denotes the tensor-by-matrix product. The choice of link function is based on the assumed distribution family of tensor entries. Common choices of link functions include identity link for Gaussian distribution, logistic link for Bernoulli distribution, and exponential link for Poisson distribution. In general, dispersion parameters can also be included in the model. Because our main focus is the tensor decomposition under the mean model, we suppress the dispersion parameter in this section for ease of presentation. 

Figure~\ref{fig:intro1}b provides a schematic illustration of our model. The features $\mX_k$ affect the distribution of tensor entries in $\tY$ through the reduced features $\mX_k\mM_k$, which are $r_k$ linear combinations of features on mode $k$. We call $\mM_k$ the ``dimension reduction matrix'' or ``tensor factors.'' The core tensor $\tC$ collects the interaction effects between reduced features across $K$ modes. We call $\tB=\tC\times\{\mM_1,\ldots,\mM_K\}$ the coefficient tensor, and $\Theta=\tB\times\{\mX_1,\cdots,\mX_K\}$ the linear predictor. By the definition of multilinear rank, the model~\eqref{eq:decomp} implies the linear predictor $\Theta$ and coefficient tensor $\tB$ are of rank-$\mr$. The conditional mean tensor $\mathbb{E}(\tY|\mX_1,\ldots,\mX_K)$ is however often high rank, due to the nonlinearity of the link function~\citep{lee2021beyond}. 

Our goal is to estimate the low-rank tensor $\tB$, or equivalently, the core tensor and factors $(\tC, \mM_1,\ldots,\mM_K)$, from our model~\eqref{eq:decomp}. { We make several remarks about model identifiability. First, the identifiability of $\tB$ requires the feature matrices $\mX_k$ are of full column rank with $p_k\leq d_k$. We impose this rank non-deficiency assumption to $\mX_k$; this is a mild condition common in literature~\citep{lock2018supervised, li2017parsimonious, li2020generalized}. In the presence of rank deficiency, we recommend to remove redundant features from $\mX_k$ before applying our method. Second, the decomposition $\tB = \tC\times\{\mM_1,\ldots,\mM_K\}$ are non-unique, as in standard tensor decomposition~\citep{kolda2009tensor}. For any invertible matrices $\mO_k\in\mathbb{R}^{r_k\times r_k}$, $\tB=\tC\times\{\mM_1,\ldots,\mM_K\}=\tC'\times\{\mM_1\mO_1,\ldots,\mM_K\mO_K\}$ are two equivalent parameterizations with $\tC'=\tC\times\{\mO^{-1}_1,\ldots,\mO^{-1}_K\}$ . To resolve this ambiguity, we impose orthonormality to $\mM_k\in\mathbb{O}(p_k,r_k)$ and assess the estimation error of $\mM_k$ using angle distance. The angle distance is invariant to orthogonal rotations due to its geometric definition. See Section~\ref{subsec:statprob} for more details. The orthonormality of $\mM_k$ is imposed purely for technical convenience. This normalization incurs no impacts in our statistical inference, but may help with numerical stability in empirical optimization~\citep{de2000multilinear, kolda2009tensor}.} Finally, the problem size is quantified by $p_k$ and $d_k$, where $p_k$ specifies the number of features and $d_k$ the number of samples at mode $k\in[K]$. Our theory treats the rank $r_k$ as known and fixed, whereas both $p_k$ and $d_k$ are allowed to increase. The adaptation to unknown rank in practice will be addressed in Section~\ref{sec:tuning}.

\subsection{Three examples}\label{sec:motivation}
We give three seemingly different examples that can all be formulated as our supervised tensor decomposition model~\eqref{eq:decomp}.
\begin{example}[Spatio-temporal growth model]
The growth curve model~\citep{gabriel1998generalised,srivastava2008models} was originally proposed as an example of bilinear model for matrix data, and we adopt its higher-order extension here. Let $\tY=\entry{y_{ijk}}\in\mathbb{R}^{d \times m\times n}$ denote the pH measurements of $d$ lakes at $m$ levels of depth and for $n$ time points. Suppose the sampled lakes belong to $q$ types, with $p$ lakes in each type. Let $\{\ell_j\}_{j\in[m]}$ denote the sampled depth levels and $\{t_k\}_{k\in[n]}$ the time points. Assume that the expected pH trend in depth is a polynomial of order at most $r$ and that the expected trend in time is a polynomial of order $s$. Then, the conditional mean model for the spatio-temporal growth can be represented as
\begin{equation}\label{eq:time}
\mathbb{E}(\tY|\mX_1,\mX_2,\mX_3)=\tC\times\{\mX_1\mM_1,\ \mX_2\mM_2,\ \mX_3\mM_3\},
\end{equation}
where $\mX_1=\text{blockdiag}\{\mathbf{1}_p,\ldots,\mathbf{1}_p\}\in \{0,1\}^{d\times q}$ is the design matrix for lake types, and
\[
\mX_2=
\begin{pmatrix}
1 & \ell_1&\cdots &\ell^{r}_1\\
1 & \ell_2&\cdots &\ell^{r}_2\\
\vdots &\vdots&\ddots&\vdots\\
1&\ell_{m}&\cdots&\ell^{r}_{m}
\end{pmatrix},\quad
\mX_3=
\begin{pmatrix}
1 & t_1&\cdots &t^{s}_1\\
1 & t_2&\cdots &t^{s}_2\\
\vdots &\vdots&\ddots&\vdots\\
1&t_{n}&\cdots&t^{s}_{n}
\end{pmatrix}
\]
are the design matrices for spatial and temporal effects, respectively, $\tC\in\mathbb{R}^{r_1\times r_2\times r_3}$ is the unknown core tensor, and $\mM_k$ are unknown dimension reduction matrices on each mode. The factors $\mX_k\mM_k$ are reduced features in the mean model~\eqref{eq:time}. The spatial-temporal model is a special case of our supervised tensor decomposition model~\eqref{eq:decomp}, with features available on each of the three modes.
\end{example}

\begin{example}[Network population model]\label{example:brain}
Network response model~\citep{rabusseau2016low} is recently developed for neuroimaging analysis. The goal is to study the relationship between brain network connectivity pattern and features of individuals. Suppose we have a sample of $n$ observations, $\{(\mY_i, \mx_i)\colon i=1,\ldots,n\}$, where for each individual $i\in[n]$, $\mY_i\in\{0,1\}^{d\times d}$ is the undirected adjacency matrix whose entries indicate presences/absences of connectivities between $d$ brain nodes, and $\mx_i\in\mathbb{R}^p$ is the individual's feature such as age, gender, cognition score, etc. The network-response model  has the conditional mean
\begin{equation}\label{eq:network}
\textup{logit}(\mathbb{E}(\mY_i|\mx_i))=\tB\times_3\mx_i, \quad \text{for }i=1,\ldots,n,
\end{equation}
where $\tB\in \mathbb{R}^{d\times d\times p}$ is a rank-$(r_1,r_1,r_2)$ coefficient tensor, and $\tB$ is assumed to be symmetric in the first two modes.  

The model~\eqref{eq:network} is a special case of our supervised tensor decomposition, with feature matrix on the last mode of the tensor. Specifically, we stack the network observations $\{\mY_i\}$ together and obtain an order-3 response tensor $\tY\in\{0,1\}^{d\times d\times n}$. Define a feature matrix $\mX=[\mx_1,\ldots,\mx_n]^T\in\mathbb{R}^{n\times p}$. Then, the model~\eqref{eq:network} has the equivalent representation of supervised tensor decomposition,
\[
\textup{logit}(\mathbb{E}(\tY|\mX))=\tC\times\{\mM,\ \mM,\ \mX\mM'\},
\]
where $\tC\in\mathbb{R}^{r_1\times r_1\times r_2}$ is the core tensor, $\mM\in\mathbb{R}^{d\times r_1}$ is the dimension reduction matrix on the first two modes, and $\mM'\in\mathbb{R}^{p\times r_2}$ is for the last mode.  \end{example}
 
 \begin{example}[Dyadic data with node attributes] Dyadic dataset consists of measurements on pairs of objects. Common examples include graphs and networks. Let $\tG=(V,E)$ denote a graph, where $V=[d]$ is the node set of the graph, and $E\subset V\times V$ is the edge set. Suppose that we also observe feature vector $\mx_i\in\mathbb{R}^p$ associated to each node $i\in V$. A probabilistic model on the graph $\tG=(V,E)$ can be described by the following matrix regression. The edge connects the two vertices $i$ and $j$ independently of other pairs, and the probability of connection is modeled as
\begin{equation}\label{eq:edge}
 \textup{logit}\left(\mathbb{P}\left((i,j)\in E\right)\right)=\mx^T_i\mB\mx_j=\langle \mB,\ \mx^T_i\mx_j\rangle,
 \end{equation}
 where $\mB\in\mathbb{R}^{p\times p}$ is a symmetric rank-$r$ matrix. The low-rankness in $\mB$ has demonstrated its success in modeling transitivity, balance, and communities in  networks~\citep{hoff2005bilinear}. We show that our supervised tensor decompostion~\eqref{eq:decomp} also incorporates the graph model as a special case. Let $\tY=\entry{y_{ij}}$ be a binary matrix where $y_{ij}=\mathds{1}_{(i,j)\in E}$. Define $\mX=[\mx_1,\ldots,\mx_n]^T\in\mathbb{R}^{n\times p}$. Then, the graph model~\eqref{eq:edge} can be expressed as
 \[
 \textup{logit}(\mathbb{E}(\mY|\mX))=\mC\times\{\mX\mM,\ \mX\mM\}, 
  \]
  where $\mC\in\mathbb{R}^{r\times r}, \mM\in\mathbb{R}^{p\times r}$ are from the singular value decomposition of $\mB=\mM\mC\mM^T$. 
  \end{example}

In the above three examples and many other studies, researchers are interested in uncovering the variation in the data tensor that can be explained by features. Our supervised tensor decomposition~\eqref{eq:decomp} allows arbitrary numbers of feature matrices. When certain mode $k$ has no side information, we set $\mX_k=\mI$ in the model~\eqref{eq:decomp}. In particular, our model~\eqref{eq:decomp} reduces to classical unsupervised tensor decomposition~\citep{de2000multilinear,hong2020generalized} when no side information is available; i.e., $\mX_k=\mI$ for all $k\in[K]$.

\section{Estimation}\label{sec:est}

\subsection{Rank-constrained MLE}\label{sec:rankM}
We develop a likelihood-based procedure to estimate $\tC$ and $\mM_k$ in~\eqref{eq:decomp}. We adopt the exponential family as a flexible framework for different data types. In a classical generalized linear model with a scalar response $y$ and feature $\mx$, the density is expressed as
\[
p(y|\mx, \boldsymbol{\beta})=c(y,\phi)\exp\left(\frac{y\theta- b(\theta)}{\phi}\right)\ \text{with}\ \theta=\boldsymbol{\beta}^T\mx,
\]
where $b(\cdot)$ is a known function, $\theta$ is the linear predictor, $\phi>0$ is the dispersion parameter, and $c(\cdot)$ is a known normalizing function. The choice of link functions depends on the data types and on the observation domain of $y$, denoted $\mathbb{Y}$. For example, the observation domain is $\mathbb{Y}=\mathbb{R}$ for continuous data, $\mathbb{Y}=\mathbb{N}$ for count data, and  $\mathbb{Y}=\{0,1\}$ for binary data. The canonical link function $f$ is chosen to be $f(\cdot)=b'(\cdot)$, the first-order derivative of $b(\cdot)$. Table~\ref{table:link} summarizes the canonical link functions for common types of distributions. 

\begin{table}[htb]
\centering
\begin{tabular}{c|cccc}
Data type &Gaussian & Poisson& Bernoulli &  \\
\hline
Domain $\mathbb{Y}$& $\mathbb{R}$&$\mathbb{N}$&$\{0,1\}$  \\
 $b(\theta)$&$\theta^2/2$& $\exp(\theta)$&$\log (1+\exp(\theta))$  \\
 link $f(\theta)$&$\theta$&$\exp(\theta)$&$(1+\exp(-\theta))^{-1}$ \\
\end{tabular}
\caption{Canonical links for common distributions.}\label{table:link}
\end{table}

In our context, we model the entries in data tensor $\tY$, conditional on linear predictor $\Theta$, as independent draws from an exponential family. Ignoring constants that do not depend on $\Theta$, the quasi log-likelihood of~\eqref{eq:decomp} is equal to Bregman distance between $\tY$ and $b'(\Theta)$:
\begin{align}\label{eq:loglikelihood}
\tL_{\tY}(\tC,\mM_1,\ldots,\mM_K)&=\langle \tY, \Theta \rangle - \sum_{i_1,\ldots,i_K} b(\theta_{i_1,\ldots,i_K}),\notag \\
 \text{where}\quad \Theta&=\tC\times\{\mX_1\mM_1,\ldots,\mX_K\mM_K\}.
\end{align}
 We propose the constrained maximum quasi-likelihood estimate (MLE),
\begin{align} \label{eq:MLE} 
(\hat \tC_{\text{MLE}}, \hat \mM_{1,\text{MLE}},\ldots,\hat \mM_{K,\text{MLE}}) =\argmax_{(\tC,\mM_1,\ldots,\mM_K)\in \tP(\mr)} \ \tL_{\tY}(\tC,\mM_1,\ldots,\mM_K),
\end{align}
where the parameter space $\tP(\mr)$ is defined by
\begin{equation}\label{eq:p}
\tP(\mr)=\left\{(\tC, \mM_1,\ldots,\mM_K) \ \Big| \ \mM_k\in\mathbb{O}(p_k,r_k)\ \text{for all }k\in[K],\  \mnormSize{}{\Theta}\leq \alpha \right\},
\end{equation}
with a large constant $\alpha>0$. Recall that $\tB=\tC\times\{\mX_1,\ldots,\mM_K\}$ by definition. Correspondingly, we estimate the coefficient tensor $\tB$ by
\[
\hat \tB_{\text{MLE}}=\hat \tC_{\text{MLE}}\times\{\hat \mM_{1,\text{MLE}},\ldots,\hat \mM_{K,\text{MLE}}\}.
\] 

The maximum norm constraint on the linear predictor $\Theta$ is a technical condition to ensures the existence (boundedness) of MLE. 
{ The condition precludes the ill-defined MLE when the optimizer of~\eqref{eq:MLE} diverges to $\pm \infty$; this phenomenon may happen in logistic regression when the Bernoulli responses $\{0,1\}$ are perfectly separable by covariates~\citep{wang2020learning}. For Gaussian models, no maximum norm constraint is needed. In Section~\ref{subsec:statprob}, we show that setting $\alpha$ to an extremely large constant does not compromise the statistical rate in quantities of interest. In practice, the unbounded search is often indistinguishable from the bounded search, since the boundary constraint $\onorm{\Theta}_{\infty} \leq \alpha$ would likely never be active. Similar techniques are commonly used in high-dimensional non-Gaussian problems~\citep{wang2020learning,han2020optimal}.}

The optimization~\eqref{eq:MLE} is a non-convex problem with possibly local optimizers. We propose an alternating optimization algorithm to {\it approximately} solve~\eqref{eq:MLE}. 
The decision variables in the objective function~\eqref{eq:MLE} consist of $K+1$ blocks of variables, one for the core tensor $\tC$ and $K$ for the factor matrices $\mM_k$. We notice that, if any $K$ out of the $K+1$ blocks of variables are known, then the optimization reduces to a simple GLM with respect to the last block of variables. This observation leads to an iterative updating scheme for one block at a time while keeping others fixed. Given an initialization $(\hat\tC^{(0)},\hat\mM_1^{(0)}, \ldots, \hat\mM_K^{(0)})$ to be described in the next paragraph, the $t$-th iterate from the algorithm is denoted $(\hat\tC^{(t)},  \hat\mM^{(t)}_1,\ldots,\hat\mM^{(t)}_K)$ for $t=1, 2,3, \ldots$ The iteration scheme is detailed in Algorithm~\ref{alg:B}. 
\begin{algorithm}[!h]
\caption{Supervised Tensor Decomposition with Side Information}\label{alg:B}
\begin{algorithmic}[1]
\INPUT Response tensor $\tY\in \mathbb{R}^{d_1\times \cdots \times d_K}$, feature matrices $\mX_k\in\mathbb{R}^{d_k\times p_k}$ for $k=1,\ldots,K$, target rank $\mr=(r_1,\ldots,r_K)$, link function $f$, initialization $(\hat\tC^{(0)},\hat \mM^{(0)}_1,\ldots,\hat\mM^{(0)}_K)$.
\For {$t=1,2,3,\ldots$}
\For { $k=1$ to $K$}
\State Obtain the factor matrix $\hat\mM^{(t)}_k\in\mathbb{R}^{p_k\times r_k}$ by a GLM with link function $f$.
\State Perform QR factorization $\hat\mM^{(t)}_k=\mQ_k\mR_k$, where $\mQ_k\in\mathbb{O}(p_k, r_k)$.
\State Update $\hat\mM^{(t)}_k\leftarrow \mQ_k$ and core tensor $\hat\tC^{(t)}\leftarrow \hat\tC^{(t)}\times_k \mR_k$.
\EndFor
\State Update the core tensor $\tC$ by solving a GLM with $\textup{vec}(\tY)$ as response, $\otimes_{k=1}^K[ \mX_k\mM_k]$ as features, and $f$ as link function. Here $\otimes$ denotes the Kronecker product of matrices. 
\EndFor
\OUTPUT factor estimate $(\hat\tC^{(t)}, \hat\mM^{(t)}_1,\ldots,\hat\mM^{(t)}_K)$ from the $t$-th iterate, and coefficient tensor estimate $\hat\tB^{(t)}=\hat\tC^{(t)}\times\{\hat\mM^{(t)}_1,\ldots, \hat\mM^{(t)}_K\}$.
\end{algorithmic}
\end{algorithm}

{ We provide two initialization schemes, one with QR-adjusted spectral initialization (warm initialization), and the other with random initialization (cold initialization). The warm initialization is an extension of unsupervised spectral initialization~\citep{zhang2018tensor} to supervised setting with multiple feature matrices.} Specifically, we project normalized data tensor $\bar \tY$ to the normalized multilinear feature space and obtain an unconstrained coefficient tensor $\hat\tB^{(0)}$. We perform a rank-$\mr$ higher-order SVD (HOSVD) on $\bar \tB$, which yields the rank-constrained $\hat\tB^{(0)}$. The desired initialization is obtained by re-normalizing $\hat\tB^{(0)}$ back to the original scales of features. The initialization scheme is described in Algorithm~\ref{alg:A}.

{ The warm initialization enjoys provable accuracy guarantees at a cost of extra technical assumptions (see Section~\ref{subsec:statprob}). The cold initialization, on the other hand, shows robust in practice but its theoretical guarantee remains an open challenge~\citep{luo2021low}. We incorporate both options in our software package to provide flexibility to practitioners.}

\begin{algorithm}[!h]
\caption{QR-adjusted spectral initialization}\label{alg:A}
\begin{algorithmic}[1]
\INPUT Response tensor $\tY\in \mathbb{R}^{d_1\times \cdots \times d_K}$, feature matrices $\mX_k\in\mathbb{R}^{d_k\times p_k}$, Tucker rank $\mr$. 
\State Normalize date tensor $\bar \tY\leftarrow \tY$ for Gaussian model, $\bar \tY\leftarrow 2\tY-1$ for Bernoulli model, and $\bar \tY\leftarrow \log(\tY+0.5)$ for Poisson model. 
\State Normalize feature matrices via QR factorization $\mX_k=\mQ_k\mR_k$ for all $k\in[K]$.
\State Obtain $\bar \tB \leftarrow \bar \tY\times\{\mQ^T_1,\ldots,\mQ^T_K\}$ by projecting $\bar \tY$ to the multilinear feature space. 
\State Obtain $\hat\tB^{(0)}\leftarrow \textup{HOSVD}(\bar \tB,\mr)$.
\State Normalize representation $\{\hat\tC^{(0)}, \hat\mM^{(0)}_1,\ldots,\hat\mM^{(0)}_K\}$ such that $\hat\tC^{(0)}\times\{ \hat\mM^{(0)}_1,\ldots,\hat\mM^{(1)}_K\} = \hat\tB^{(0)}\times\{ \mR^{-1}_1,\ldots,\mR^{-1}_K\}$ and $\hat\mM^{(0)}_k\in\mathbb{O}(p,r)$ for all $k\in[K]$.
\OUTPUT Core tensor $\hat\tC^{(0)}$ and factors $\hat\mM^{(0)}_k$ for all $k\in[K]$.
\end{algorithmic}
\end{algorithm}

\subsection{Statistical accuracy}\label{subsec:statprob}
{ This section presents the accuracy guarantees for both global and local optimizers of~\eqref{eq:MLE}. We first provide the statistical accuracy for the global MLE~\eqref{eq:MLE}. Then, we provide the convergence rate for the local optimizer from Algorithm~\ref{alg:B} with warm initialization. The rate reveals an interesting interplay between statistical and computational efficiency. We show that a polynomial number of iterations suffices to reach the desired accuracy under certain assumptions.} The empirical performance for cold initialization is also investigated. 

For cleaner exposition, we present the results for balanced setting in this section, i.e., $p_1=\cdots=p_K=p$, $r_1=\cdots=r_K=r$, and $d_1=\cdots=d_K=d$. The general setting follows exactly the same framework and incurs only notational complexity. We are particularly interested in the high-dimensional regime in which both $d$ and $p$ grows while $p\leq d$. { The requirement $p\leq d$ is necessary to ensure rank non-deficiency of  feature matrices $\mX_k$.} The classical MLE theory is not directly applicable, because the number of unknown parameters grows with the size of data tensor. We leverage the recent development in random tensor theory and high-dimensional statistics to establish the error bounds of the estimation.

\begin{assumption}\label{ass}We make the following assumptions:
\begin{enumerate}[noitemsep,topsep=0pt]
\item [A1.] There exist two positive constants $c_1, c_2>0$ such that $c_1\leq \sigma_{\min}(\mX_k)\leq  \sigma_{\max}(\mX_k)\leq c_2$ for all $k\in[K]$. Here $\sigma_{\min}(\cdot)$ and $\sigma_{\max}(\cdot)$ denote the smallest and largest matrix singular values.
\item [A1'.] The feature matrices $\mX_k$ are Gaussian designs with i.i.d.\ $N(0,1)$ entries.
\item [A2.] There exist two positive constants $L, U>0$, such that $\min_{|\theta|\leq \alpha}b''(\theta)\geq \phi L$ and $\sup_{\theta\in \mathbb{R}}b''(\theta)\leq \phi U$. Here, $\alpha$ is the upper bound of the linear predictor in~\eqref{eq:MLE}, and $b''(\cdot)$ denotes the second-order derivative.
\end{enumerate}
\end{assumption}
The assumptions are fairly mild. Assumptions A1 and A1' consider two separate scenarios about feature matrices. Assumption A1 is applicable when feature matrix is asymptotically non-singular and has bounded spectral norm, whereas Assumption A1' imposes the commonly-used Gaussian design~\citep{raskutti2019convex}.  The Assumption 2 is essentially imposed to the response variance because of the identity $\text{Var}(y|\theta)=\phi b''(\theta)$~\citep{mccullagh1989generalized}. The lower bound  ensures the non-degeneracy of the variance in the feasible domain of $\theta$, whereas the upper bound ensures the finiteness of the variance in the entire family. In fact, except for Poisson responses, most members in the exponential family, e.g., Gaussian, Bernoulli, and binomial responses, satisfy this condition.

\subsubsection{Statistical accuracy for global optimizers}\label{sec:global}
We need some extra notation to state the results in full generality. Recall that the factor matrices $\mM_k$ are identifiable only up to orthogonal rotations. Therefore, we choose to use angle distance to assess the estimation accuracy of $\mM_k$. For any two column-orthonormal matrices $\mA,\mB\in\mathbb{O}(d,r)$ of same dimension, the angle distance is defined as
\[
\sin \Theta(\mA,\mB)=\max\left\{ \frac{\langle \mx, \my\rangle}{\vnormSize{}{\mx}\vnormSize{}{\my}}\colon \ \mx\in \textup{Span}(\mA),\ \my\in \textup{Span}(\mB^{\perp})\right\},
\]
where $\text{Span}(\cdot)$ represents the column space of the matrix. We use the superscript ``true'' to denote the true parameters from generic decision variables in optimization. For instance, $\trueB$ denotes the true coefficient tensor, whereas $\tB$ denotes a decision variable in~\eqref{eq:loglikelihood}. 

Define the signal level $\lambda$ as the minimal singular value of the unfolded matrices obtained from $\trueB$,  
\[
\lambda=\min_{k\in[K]}\sigma_{r}(\text{Unfold}_k(\tB_{\text{true}})).
\]
Intuitively, $\lambda$ quantifies the level of rank non-degeneracy for the true coefficient tensor $\trueB$. 

\begin{thm}[Statistical rate for global optimizers]\label{thm:MLE}Consider generalized tensor models with multiple feature matrices. Under Assumptions A1 and A2 with scaled feature matrices $\bar \mX_k= \sqrt{d}\mX_k$, or Assumptions A1' and A2 with original feature matrices, we have
\begin{equation}\label{eq:bound}
\max_{k\in[K]}\sin^2\Theta(\trueM, \hat \mM_{k,\text{MLE}})\lesssim {\phi (r^K+Kpr)\over \lambda^2 d^K}, \quad \FnormSize{}{\tB_{\text{true}}-\hat \tB_{\text{MLE}} }^2\lesssim {\phi (r^K+Kpr) \over d^K},
\end{equation}
with probability at least $1-\exp(-p)$.
\end{thm}
Theorem~\ref{thm:MLE} establishes the statistical convergence for the global MLE~\eqref{eq:MLE}. The result in~\eqref{eq:bound} implies that the estimation has a convergence rate $\tO(Kp/d^{K})$ as $(p,d)\to \infty$. { This rate agrees with intuition, since in our setting, the number of parameters with $K$ feature matrices is of order $\tO(Kp)$, whereas the number of tensor entries $\tO(d^K)$ corresponds to the total sample size. Because $p\leq d$, our rate is faster than $\tO(d^{-(K-1)})$ obtained by tensor decomposition without features~\citep{wang2020learning}.} 

Inspection of our proof in Section~\ref{subsec:proof_MLE} shows that the desired convergence rate holds not only for the MLE, but also for all local optimizers satisfying $\tL_{\tY}(\tC, \mM_1,\ldots,\mM_K)\geq \tL_{\tY} (\tC_{\textup{true}},\mM_{1,\textup{true}},\ldots,\mM_{K,\textup{true}})$. The observation indicates the global optimality is not necessarily a serious concern in our context, as long as the convergent objective is large enough. In next section, we will provide the statistical accuracy for \emph{local} optimizer with provable convergence guarantee, at a cost of extra signal requirement.

\subsubsection{Empirical accuracy for local optimizers}\label{sec:local}

The optimization~\eqref{eq:MLE} is a non-convex problem due to the low-rank constraint in the feasible set $\tP$. { Under mild conditions, our warm initialization enjoys stable performance, and the subsequent iterations further improve the accuracy via linear convergence; i.e.\ sequence of iterates generated by Algorithm~\ref{alg:B} converges to optimal solutions at a linear rate. 

\begin{prop}[Polynomial-time angle estimation]\label{lem:ini} Consider Gaussian tensor models with $b(\theta)=\theta^2/2$ in the objective function~\eqref{eq:loglikelihood}. Suppose the signal-to-noise ratio $\lambda^2/\phi \geq C p^{K/2}d^{-K}$ for some sufficiently large universal constant $C>0$. 
Under Assumption A1 with scaled feature matrices $\bar \mX_k=\sqrt{d}\mX_k$, or Assumption A1' with original feature matrices, the outputs from initialization Algorithm~\ref{alg:A} and iteration Algorithm~\ref{alg:B} satisfy the following two properties.
\begin{enumerate}[label=(\alph*)]
\item With probability at least $1-\exp(-p)$. 
\begin{equation}\label{eq:warm}
\max_{k\in[K]}\sin^2\Theta(\trueM,\hat \mM_k^{(0)})\leq {1\over 4}. 
\end{equation}
\item Let $t=1,2,3,\ldots,$ denote the iteration. There exists a contraction parameter $\rho\in(0,1)$, such that, with probability at least $1-\exp(-p)$, 
\begin{equation}\label{eq:local}
\max_{k\in[K]}\textup{sin}^2 \Theta(\trueM, \hat \mM^{(t)}_k) \lesssim \KeepStyleUnderBrace{{\phi p \over \lambda^2 d^K}}_{\text{statistical error}}+\KeepStyleUnderBrace{\rho^{t}\max_{k\in[K]}\textup{sin} \Theta^2(\trueM, \hat \mM_k^{(0)})}_{\text{algorithmic error}}.
\end{equation}
\end{enumerate}
\end{prop}

Proposition~\ref{lem:ini} provides the estimation errors for algorithm outputs at initialization and at each of the subsequent iterations. The initialization bound~\eqref{eq:warm} demonstrates the stability of warm initialization under a mild SNR requirement $\lambda^2/\phi \gtrsim p^{K/2}d^{-K}$. We can think of $d$ as the sample size while $p$ the number of parameters at mode $K$. This threshold is less stringent than $d^{K/2}$ required for unsupervised tensor decomposition features~\citep{han2020optimal,zhang2018tensor}. The condition confirms that a higher sample size mitigates the required signal level. The iteration bound~\eqref{eq:local} consists of two terms: the first term is the statistical error, and the second is the algorithmic error. The algorithmic error decays exponentially with the number of iterations, whereas the statistical error remains the same as $t$ grows. The statistical error is unavoidable and also appears in the global MLE; see Theorem~\ref{thm:MLE}.

As a direct consequence, we find the optimal iteration $t$ after which the algorithmic error is negligible compared to statistical error. 
\begin{thm}[Statistical rate for local optimizers]\label{thm:local}Consider the same condition as in Proposition~\ref{lem:ini} and the outputs by combining algorithms 1 and 2. There exists a constant $C>0$, such that, after $t\gtrsim K\log_{1/\rho}p$ iterations, our algorithm outputs satisfies
\[
\max_{k\in[K]}\textup{sin}^2 \Theta(\trueM, \hat \mM^{(t)}_k) \lesssim {\phi p\over \lambda^2d^K}, \quad \FnormSize{}{\trueB- \hat \tB^{(t)}}^2 \lesssim {\phi(r^K+Kpr)\over d^K}.
\]
\end{thm}
}

\begin{figure}[t]
\centering
\includegraphics[width=15cm]{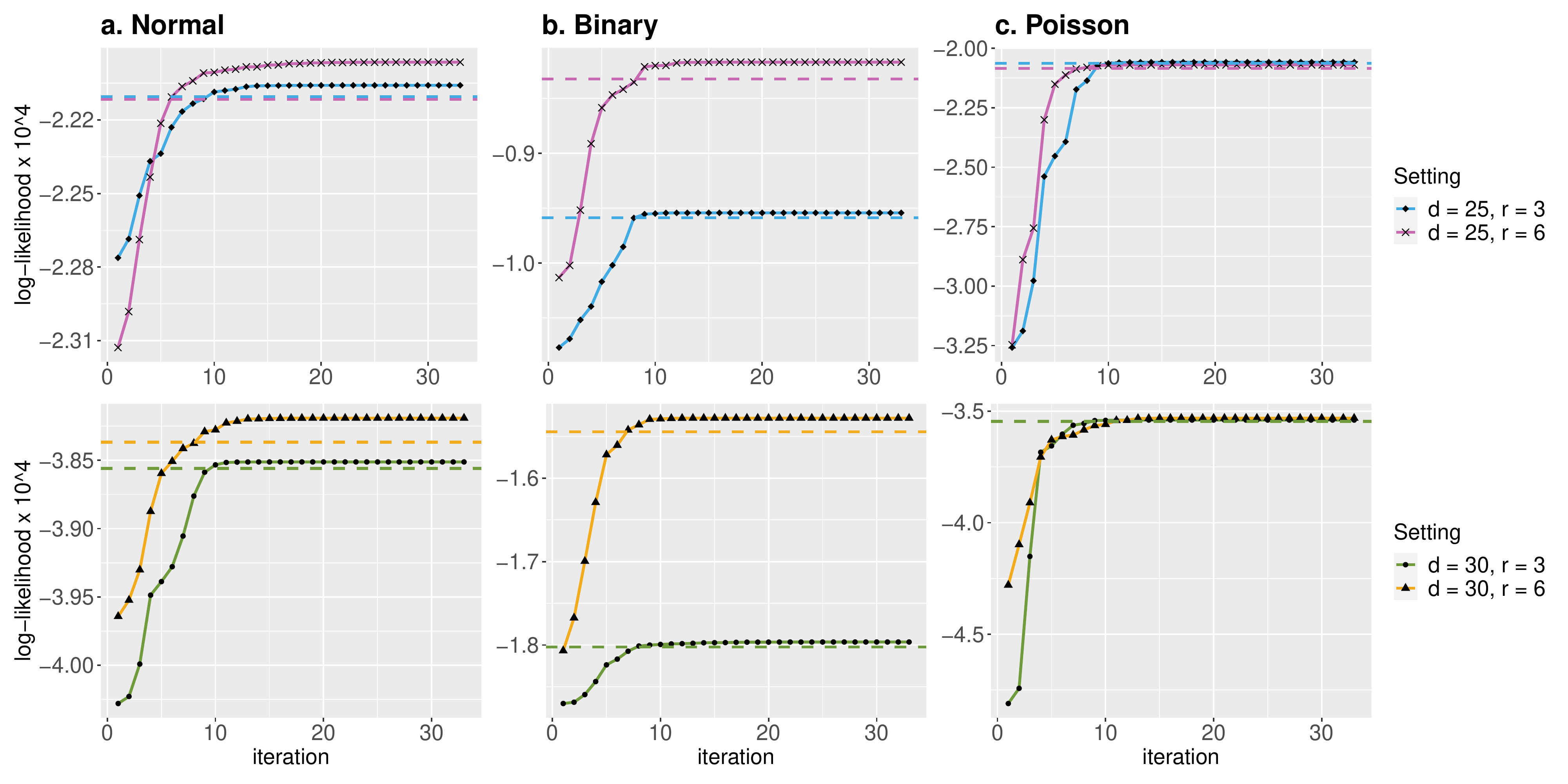}
\caption{Trajectory of the objective function with various dimension $d$ and rank $r$ under (a) Gaussian (b) Bernoulli (c) Poisson models. The dashed line represents the objective value at true parameters. }\label{fig:loglike}
\vspace{-.2cm}
\end{figure}

In practice, the signal level $\lambda$ is unknown, so the assumption in Theorem~\ref{thm:local} is challenging to verify in practice. We supply the theory by providing an alternative scheme -- random initialization -- and investigate its empirical performance. Figure~\ref{fig:loglike} shows the trajectories of objective function for order-3 tensors based on model~\eqref{eq:decomp}, where $ d \in\{25,30\}$, $p = 0.4d$, $ r\in\{3,6\}$ at all three modes. We consider data tensors with Gaussian, Bernoulli, and Poisson entries. Under all combinations of the dimension $d$, rank $r$, and type of the entries, Algorithm~\ref{alg:B} converges quickly in a few iterations upon random initialization, and the objective values at convergent points are close to or larger than the value at true parameters. In the experiment we conduct, we find little difference in the final estimation errors between the two initialization schemes. Random initialization appears good enough for Algorithm~\ref{alg:B} to find a convergent point with desired statistical guarantees. In practice, we recommend to run both warm and cold initializations, and choose the one with better convergent objective values.

We conclude this section by revisiting the three examples mentioned in Section~\ref{sec:model}.

\begin{customexample}{1}[Spatio-temporal growth model] The estimated type-by-time-by-space coefficient tensor converges at the rate $\tO\left((p+r+s)/(dmn)\right)$ with $(p,r,s)\leq (d,m,n)$. The estimation achieves consistency as the dimension grows along either of the three modes.
\end{customexample}

\begin{customexample}{2} [Network population model] The estimated node-by-node-by-feature tensor converges at the rate $\tO\left((2d+p)/(d^2n)\right)$ with $p\leq n$. The estimation achieves consistency as the number of individuals or the number of nodes grows. 
\end{customexample}

\begin{customexample}{3} [Dyadic data with node attributes] The estimated feature-by-feature matrix converges at the rate $\tO\left(p/d^2\right)$ with $p\leq d$. Again, our estimation achieves consistency as the number of nodes grows. 
\end{customexample}

\subsection{Rank selection and computational complexity}\label{sec:tuning}
Our algorithm assumes the rank $\mr$ is given. In practice, the rank is often unknown and must be determined from the data. We propose to use Bayesian information criterion (BIC) and choose the rank that minimizes BIC, where
\begin{equation}\label{eq:BIC}
\textup{BIC}(\mr)=-2\tL_{\tY}(\hat \tC, \hat \mM_1,\ldots,\hat \mM_K)+p_e(\mr)\log (\prod\nolimits_k d_k).
\end{equation}
Here, $p_e(\mr)\stackrel{\textup{def}}{=}\sum_k (p_k-r_k)r_k+\prod_k r_k$ is the effective number of parameters in the model. We choose $\hat \mr$ that minimizes $\textup{BIC}(\mr)$ via grid search. Our choice of BIC aims to balance between the goodness-of-fit for the data and the degree of freedom in the population model. We evaluate the empirical performance of BIC in Section~\ref{sec:simulation}.  

The computational complexity of our Algorithm is $\tO\left(d \sum_k p^3_k\right)$ for each iteration, where $d=\prod_k d_k$ is the total size of the data tensor. The update of $K$ factor matrices is $\tO(d\sum_k r^3_k p_k^3)$ via standard GLM routines. Furthermore, we demonstrate that, under certain SNR conditions, a polynomial number of iterations suffices to reach the desired statistical accuracy. Therefore, the total computational cost is polynomial in $p$ and $d$.

\section{Connection to other tensor regression methods}\label{sec:connection}
{ We compare our supervised tensor decomposition (\textbf{STD}) with recent 12 tensor methods in the literature. Table~\ref{table:comp_table} summarizes these methods with their properties from four aspects: i) model specification, ii) number of feature matrices allowed, (iii) capability of addressing non-Gaussian response, and (iv) capability of addressing non-independent noise. The four closet methods to our are {\bf SupCP}~\citep{lock2018supervised}, {\bf Envelope}~\citep{li2017parsimonious}, {\bf mRRR}~\citep{luo2018leveraging} and {\bf GLSNet}~\citep{zhang2018network}; these methods all relate a data tensor to feature matrices with low-rank structure on the coefficients. As seen from the table, our method is the only one that allows multiple feature matrices among the five. {\bf Envelope} and {\bf SupCP} are developed for Gaussian data, and the Gaussianity facilities flexible extension to non-independent noise. In particular, {\bf Envelope} allows noise correlation in Kronecker structured form, whereas {\bf SupCP} allows noise correlation implicitly through decomposing the latent factors into fixed effects (related to features) and random effects (unrelated to features). On the other hand, the other three methods ({\bf mRRR}, {\bf GLSNet} and {\bf STD}) are developed for exponential family distribution with possibly non-additive noise. The generality makes the full modeling of noise correlation computationally challenging. We will compare the numerical performance of these methods in Section~\ref{sec:simulation}.

\afterpage{
\begin{landscape}
 \renewcommand{\arraystretch}{1.5} 
\begin{table}[h!]
\centering
 
\resizebox{\columnwidth}{!}{
\begin{tabular}{c||c|ccc}
\hline
Method & Model& No.\ of features & non-Gaussianity & Non-independence \\
\hline
	STD (Ours) & $\bbE\tY = f(\tB \times \{\mX_1,\mX_2,\mX_3\}),\ \tB = \tC \times \{\mM_1,\mM_2,\mM_3\}$ &$3$& $\surd$& $\times$\\
	GCP, CP-ARP, CORALS & $ \bbE\tY = f(\entry{ \mA_1,\mA_2,\mA_3})$ &0& $\surd$& $\times$\\
         DCOT & $\bbE\tY = f((\tC_1+\tC_2) \times \{\mM_1, \mM_2, \mM_3 \})$ &0& $\surd$ & $\times$\\
         LRT, CRT & $y_{n} = \langle \tB,\ \tX_{n} \rangle + \epsilon_{n}$, various structure on $\tB$&0& $\times$& $\times$\\
	STAR & $y_n = \sum_m \langle \tB_m, \tF_m(\tX_{ijk})\rangle + \epsilon_n$, sparse-CP $\tB_m$ &0& $\times$ & $\times$\\
	SupCP &   $\tY = \entry{ \mA_1,\mA_2,\mA_3} + \tE,\  \mA_1 = \mX \mB +\tE',\ \tE\perp \tE'$ &1& $\times$& $\surd$\\
	mRRR &  $\bbE \mY =  f(\mX \mB)$, low-rank $\mB$ &1& $\surd$& $\times$ \\
	Envelope & $\tY = \tB \times_3 \mX + \tE,\  \tB = \tC \times \{ \mM_1, \mM_2, \mI \},\  \tE \sim \mathcal{TN}(\mSigma_1, \mSigma_2,\mI)$ &1& $\times$ & $\surd$\\
	GLSNet& $\bbE\tY = f({\bf 1} \otimes {\bf \Theta} + \tB \times_3 \mX)$, low-rank ${\bf \Theta}$, sparse $\tB$&1& $\surd$& $\times$\\
	STORE & $\tY = \tB \times_3 \mX + \tE$,  sparse-CP $\tB$ &1& $\times$& $\times$ \\
	\hline
\end{tabular}
}
\caption{\footnotesize Comparison of tensor regression/factorization methods. We focus on order-3 tensors for illustration. Calligraphic letters denote tensors, bold capital letters denote matrices, and little letters denote scalars. The dimension of tensors and matrices can be identified from the contexts.\\
\vspace{-0.1cm}\\
- Data: tensor response $\tY$, feature matrices $\mX,\mX_k$, predictor tensor $\tX_n$, scalar response $y_n$, sample index $n$, tensor mode $k=1,2,3$.  \\
- Parameter: Tuckor factors $\mM_k$, CP factors $\mA_k$, CP decomposition $\entry{\mA_1, \mA_2, \mA_3}$, coefficient tensor and matrix $\tB, \tB_m, {\bf \Theta}, \mB$.\\
- Function: a known link function $f(\cdot)$, a known basis function $\tF_m(\cdot)$. \\
- Noise: Gaussian tensor with i.i.d.\ entries $\tE, \tE'$, Gaussian tensor with Kronecker covariance $\tE \sim \mathcal{TN}(\mSigma_1, \mSigma_2, \mI)$, meaning $\text{Cov}(\text{vec}(\tE))=\mSigma_1\otimes\mSigma_2\otimes \mI$.\\
\vspace{-0.1cm}\\
- GCP: Generalized canonical polyadic tensor decomposition~\citep{hong2020generalized};\\
- CP-APR: CP alternating Poisson regression~\citep{chi2012tensors};\\
- CORALS: Generalized co-clustering method~\citep{li2020generalized};\\
- DCOT: Double core tensor decomposition~\citep{tarzanagh2019regularized};\\
- SupCP: Supervised PARAFAC/CANDECOMP factorization~\citep{lock2018supervised};\\
- mRRR: Mixed-response reduced-rank regression~\citep{luo2018leveraging};\\
- Envelope: Parsimonious tensor response regression~\citep{li2017parsimonious};\\
- GLSNet: Generalized connectivity matrix response regression~\citep{zhang2018network};\\
- STORE: Sparse tensor response regression~\citep{sun2017store};\\
- LTR: Low-rank tensor regression~\citep{han2020optimal};\\
- CRT: Convex regularized multi-response tensor regression~\citep{raskutti2019convex};\\
- STAR: Sparse tensor additive regression~\citep{hao2019sparse}.
}\label{table:comp_table}
\end{table}
\end{landscape}
}
}

Our model also has a close connection to higher-order interaction model~\citep{hao2020sparse} and tensor-to-tensor regression~\citep{lock2018tensor}. Model~\eqref{eq:decomp} can be viewed as a regression model with across-mode interactions in the reduced feature space. We take an order-3 tensor under the Gaussian model for illustration. Let $\mX,\mZ,\mW$ denote the feature matrix on mode $k=1, 2, 3$, respectively. Suppose that each mode has two-dimensional reduced features, denoted $\mM_1\mX=[\mx_1,\mx_2]$, $\mM_2\mZ=[\mz_1,\mz_2]$, $\mM_3\mW=[\mw_1,\mw_2]$. Here $\mx_1,\mx_2,\ldots,\mw_1, \mw_2$ are column vectors. Then the model~\eqref{eq:decomp} is equivalent to a regression model with across-mode interactions
\begin{equation}
\mathbb{E}(y_{ijk}|\mX,\mZ,\mW)=c_{111}\mx_{1i}\mz_{1j}\mw_{1k}+c_{121}\mx_{i1}\mz_{j2}\mw_{k1}+\cdots+c_{221}\mx_{2i}\mz_{2j}\mw_{1k}+c_{222}\mx_{2i}\mz_{2j}\mw_{2k},
\end{equation}
where $\entry{c_{ijk}} \in \mathbb{R}^{2\times 2\times 3}$ are unknown interaction effects, $\mx_{1i}$ denotes the $i$-th entry in the feature vector $\mx_1$, and similar notations apply to other features. Note that lower-order interactions are naturally incorporated if we include an intercept column in the reduced feature matrices. The above example shows the connection of our supervised tensor decomposition to multivariate regressions.

\section{Numerical experiments}\label{sec:simulation}
We evaluate the empirical performance of our supervised tensor decomposition (\textbf{STD}) through simulations. We consider order-3 tensors with a range of distribution types. Unless otherwise specified, the conditional mean tensor is generated form model~\eqref{eq:decomp}, where the core tensor entries are i.i.d.\ drawn from Uniform[-1,1], the factor matrix $\mM_k$ is uniformly sampled with respect to Haar measure from matrices with orthonormal columns. The feature matrix $\mX_k$ is either an identity matrix (i.e.,\ no feature  available) or Gaussian random matrix with i.i.d.\ entries from $N(0,1)$. The linear predictor $\Theta=\tC\times\{\mM_1\mX_1,\mM_2\mX_2,\mM_3\mX_3\}$ is scaled such that $\mnormSize{}{\Theta}=1$. Conditional on the linear predictor $\Theta=\entry{\theta_{ijk}}$, the entries in the tensor $\tY=\entry{y_{ijk}}$ are drawn independently according to three probabilistic models:

\begin{enumerate}[noitemsep,topsep=0pt]
\item[(a)] Gaussian model: continuous tensor entries $y_{ijk}\sim N\left(\alpha \theta_{ijk}, 1\right)$.
\item[(b)] Poisson model: count tensor entries $y_{ijk}\sim\textup{Poisson}\left(e^{\alpha \theta_{ijk}}\right)$.
\item[(c)] Bernoulli model: binary tensor entries $y_{ijk}\sim \textup{Bernoulli}\left( \frac{e^{\alpha \theta_{ijk}}}{1+e^{\alpha \theta_{ijk}}}\right)$.
\end{enumerate}
Here $\alpha>0$ controls the magnitude of the effect size, which is also the maximum norm of coefficient tensor as in~\eqref{eq:p}.
In each experiment, we report the summary statistics averaged across $30$ simulation replications. 

\subsection{Finite-sample performance}
The first experiment assesses the selection accuracy of our BIC criterion~\eqref{eq:BIC}. We consider the balanced situation where $d_k=d$, $p_k=0.4d_k$ for $k=1,2,3$. We set $\alpha=4$ and consider various combinations of dimension $d$ and rank $\mr=(r_1,r_2,r_3)$. For each combination, we minimize BIC using a grid search from $(r_1-3,r_2-3,r_3-3)$ to $(r_1+3,r_2+3,r_3+3)$. We remove invalid rank such as $r^2_{\max} \geq \prod_{k=1}^3 r_k$ and use parallel search to reduce the computational cost. Table~\ref{tab:rank} reports the selected rank averaged over $n_{\textup{sim}}=30$ replicates. We find that in the high-rank setting with $d=20$, the selected rank slightly underestimates the true rank, and the accuracy immediately improves when either the dimension increases to $d = 40$ or the rank reduces to $\mr = (3,3,3)$. This agrees with our expectation, because in the tensor decomposition, the sample size is related to the number of tensor entries. A larger $d$ implies a larger sample size, so the BIC selection becomes more accurate. 

\begin{table}[tb]
\resizebox{\textwidth}{!}{%
\centering
\begin{tabular}{c|cc|cc}
True rank $\mr$& $d=20$  (Gaussian) &$d=40$ (Gaussian) &$d=20$ (Poisson) &$d=40$ (Poisson)\\
\hline
$(3,\ 3,\ 3)$&$({\bf 3.0},\ {\bf 3.0},\ {\bf 3.0})$&$({\bf 3.0},\ {\bf 3.0},\ {\bf 3.0})$& $({\bf 3.0},\ {\bf 3.0},\ {\bf 3.0})$ & $({\bf 3.0},\ {\bf 3.0},\ {\bf 3.0})$\\
$(4,\ 4,\ 6)$&$(3.0,\ 3.0,\ {\bf 4.6})$&$({\bf 4.0},\ {\bf 4.0},\ {\bf 5.3})$&$(3.0,\ 3.0,\ {\bf 5.3})$&$({\bf 4.0},\ {\bf 4.0},\ {\bf 5.6})$\\
$(6,\ 8,\ 8)$&$(5.0,\ 5.0,\ 5.0)$&$({\bf 6.0},\ {\bf 8.0},\ {\bf 8.0})$&$(5.0,\ 5.0,\ 6.7)$&$({\bf 6.0},\ {\bf 8.0},\ {\bf 8.0})$\\
\end{tabular}
}
\caption{Rank selection via BIC. The estimated ranks are averaged across 30 simulation. Bold number indicates the ground truth is within two standard deviations of the estimate.}\label{tab:rank}
\end{table}

\begin{figure}[!h]
\centering
\includegraphics[width=14cm]{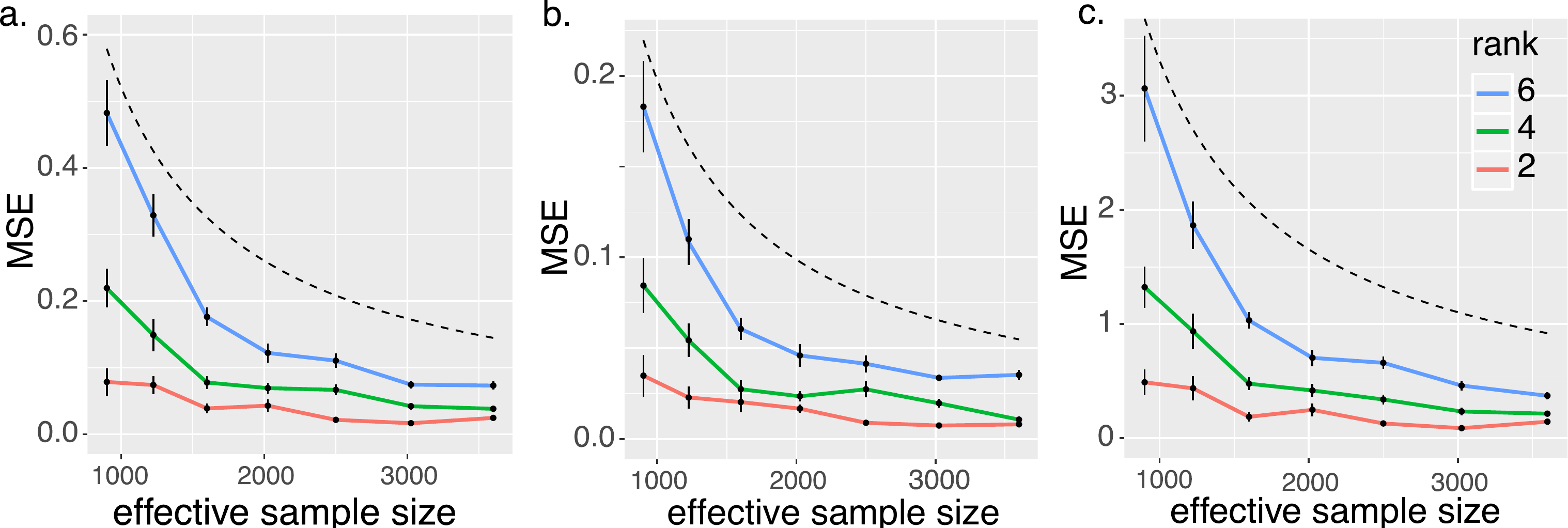}
\caption{Estimation error against effective sample size. The three panels plot the MSE when the response tensors are generated from (i) Gaussian (ii) Poisson and (iii) Bernoulli models. The dashed curves correspond to $\tO({1/d^2})$.}\label{fig:dim}
\vspace{-.1cm}
\end{figure}

The second experiment evaluates the accuracy when features are available on all modes. We set $\alpha=10, d_k=d, p_k=0.4d_k, r_k=r\in\{2,4,6\}$ and increase $d$ from 30 to 60. Our theoretical analysis suggests that $\hat \tB$ has a convergence rate $\tO(d^{-2})$ in this setting. Figure~\ref{fig:dim} plots the mean squared error (MSE) $\FnormSize{}{\hat \tB-\tB_{\text{true}}}^2$ versus the effective sample size $d^2$ under three different distribution models. We find that the empirical MSE decreases roughly at the rate of $1/d^2$, which is consistent with our theoretical results. We also observe that, tensors with higher rank tend to yield higher estimation errors, as reflected by the upward shift of the curves as $r$ increases. Indeed, a larger $r$ implies a higher model complexity and thus greater difficulty in the estimation.

\subsection{Comparison with other tensor methods}
We compare our supervised tensor decomposition with three other tensor methods:

{ \begin{itemize}[noitemsep,topsep=0pt]
\item Supervised PARAFAC/CANDECOMP factorization (\textbf{SupCP}, \citep{lock2018supervised}).
    \item Parsimonious tensor response regression (\textbf{Envelope}, \citep{li2017parsimonious});
    \item Mixed-response reduced-rank regression (\textbf{mRRR}, \citep{luo2018leveraging});
    \item Generalized connectivity matrix response regression (\textbf{GLSNet}, \citep{zhang2018network});
\end{itemize}
}

{ These four methods are the closest methods to ours, in that they all relate a data tensor to feature matrices with low-rank structure on the coefficients.} We consider Gaussian and Bernoulli tensors in experiments. For methods not applicable for Bernoulli data ({\bf SupCP} and {\bf Envelope}), we provide the algorithm $\{-1,1\}$-valued tensors as inputs. Because {\bf mRRR} allows matrix response only, we provide the algorithm the unfolded matrix of response tensor as inputs. We measure the accuracy using the response error defined as $1-\text{Cor}(\hat \tY, f(\trueT))$, where $\hat \tY$ is the fitted tensor from each method, and $f(\trueT)$ is the true conditional mean of the tensor. Note that the response error is a scale-insensitive metric; a smaller error implies a better fit of the model.

{ The comparison is assessed from three aspects: (i) benefit of incorporating features from multiple modes; (ii) prediction error with respect to sample size; (iii) robustness of model misspecification.} We use similar simulation setups as in our first experiment in last section. We consider rank $\mr=(3,3,3)$ (low) vs.\ $(4,5,6)$ (high), effect size $\alpha = 3$ (low) vs.\ $6$ (high), dimension $d$ ranging from 20 to 100 for modes with features, and $d = 20$ for modes without features. The method \textbf{Envelope} and \textbf{mRRR} require the tensor rank as inputs, respectively. For fairness, we provide both algorithms the true rank. The methods \textbf{SupCP} and \textbf{GLSNet} use different notions of model rank, and \textbf{GLSNet} takes sparsity as an input. We use a grid search to set the hyperparameters in \textbf{SupCP} and \textbf{GLSNet} that give the best performance.

\begin{figure}[t]
\centering
\includegraphics[width=12cm]{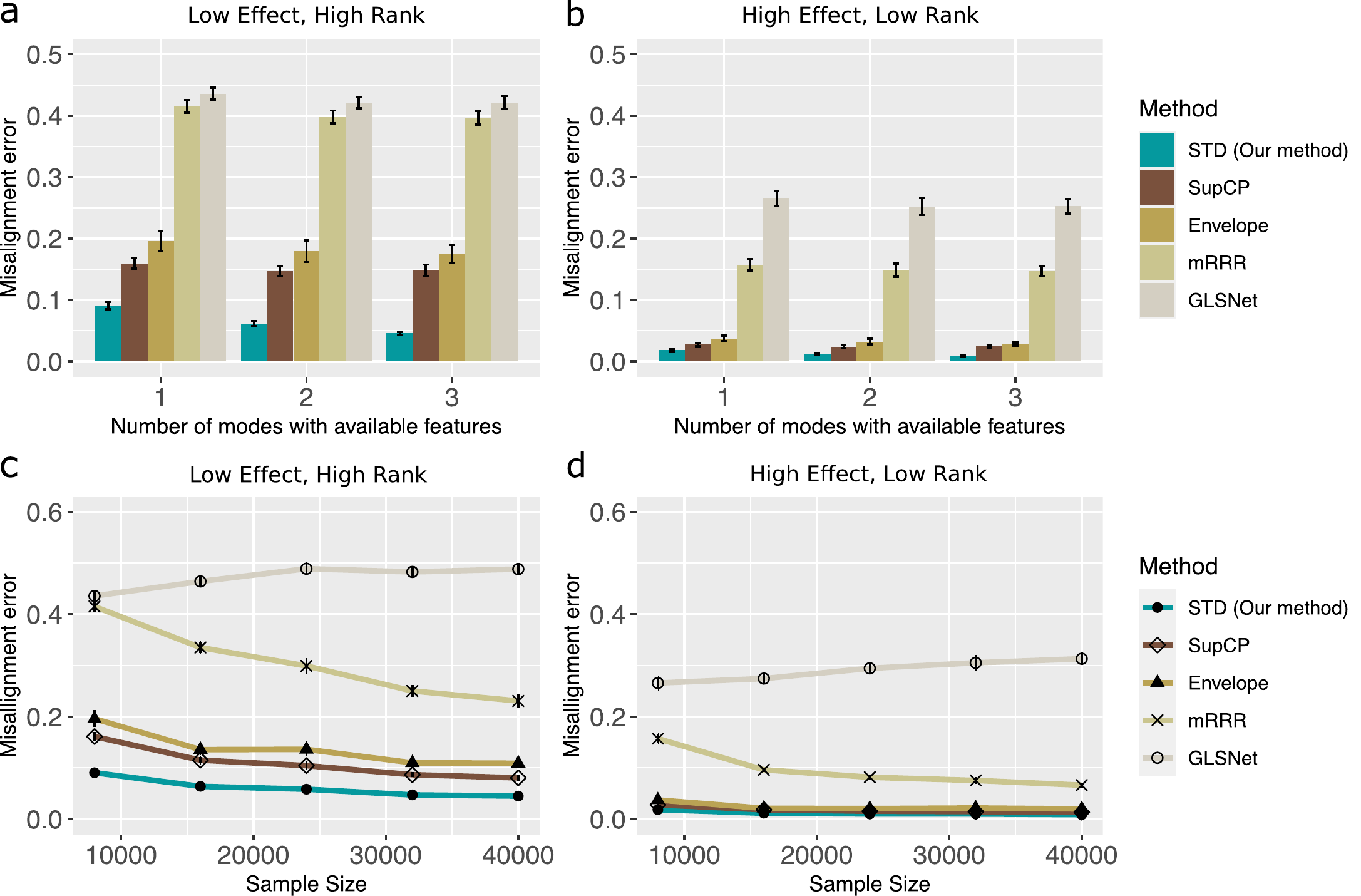} 
\caption{Comparison between tensor methods with Gaussian data. Panels (a) and (b) plot estimation error versus the number of modes with available features. Panels (c) and (d) plot ME versus the effective sample size $d^2$.
We consider rank $\mr=(3,3,3)$ (low), $\mr=(4,5,6)$ (high), and effect size $\alpha =3 $ (low), $\alpha=6$ (high).}~\label{fig:comp}
\vspace{-.5cm}
\end{figure}

{ Figure~\ref{fig:comp}a-b shows the impact of features to estimation error. We see that our \textbf{STD} outperforms others, especially in the low-effect high-rank setting. As the number of informative modes increases, the \textbf{STD} exhibits a reduction in error whereas others remain unchanged. The accuracy gain in Figure~\ref{fig:comp} demonstrates the benefit of incorporating informative features from multiple modes. In addition, we find that the relative performance among the competing methods reveals the benefits of low-rankness. 
The second best method is \textbf{SupCP} which imposes low-rankness on three modes; the next one is \textbf{Envelope} which imposes low-rankness on two modes; the less favorable one is \textbf{mRRR} which imposes low-rank structure on one mode only; the worst one is \textbf{GLSNet} which imposes sparsity but no low-rankness on the feature effects. 

Figure~\ref{fig:comp}c-d compares the prediction error with respect to effective sample size $d^2$. For fair comparison, we consider the setting with feature matrix on one mode only. We find that our \textbf{STD} method has similar performance as \textbf{Envelope} and \textbf{SupCP} in the high-effect low-rank regime, whereas the improvement becomes more pronounced in the low-effect high-rank regime. The latter setting is notably harder, and our \textbf{STD} method shows advantage in addressing this challenge. Among other methods, \textbf{Envelope}, \textbf{SupCP}, and \textbf{mRRR} show decreasing errors as $d$ increases, implying the benefits of low-rankness methods. In contrast, \textbf{GLSNet} suffers from non-decreasing error and indicates the poor fit of sparsity methods in addressing low-rank data. 

\begin{figure}[t]
\centering
\includegraphics[width=12cm]{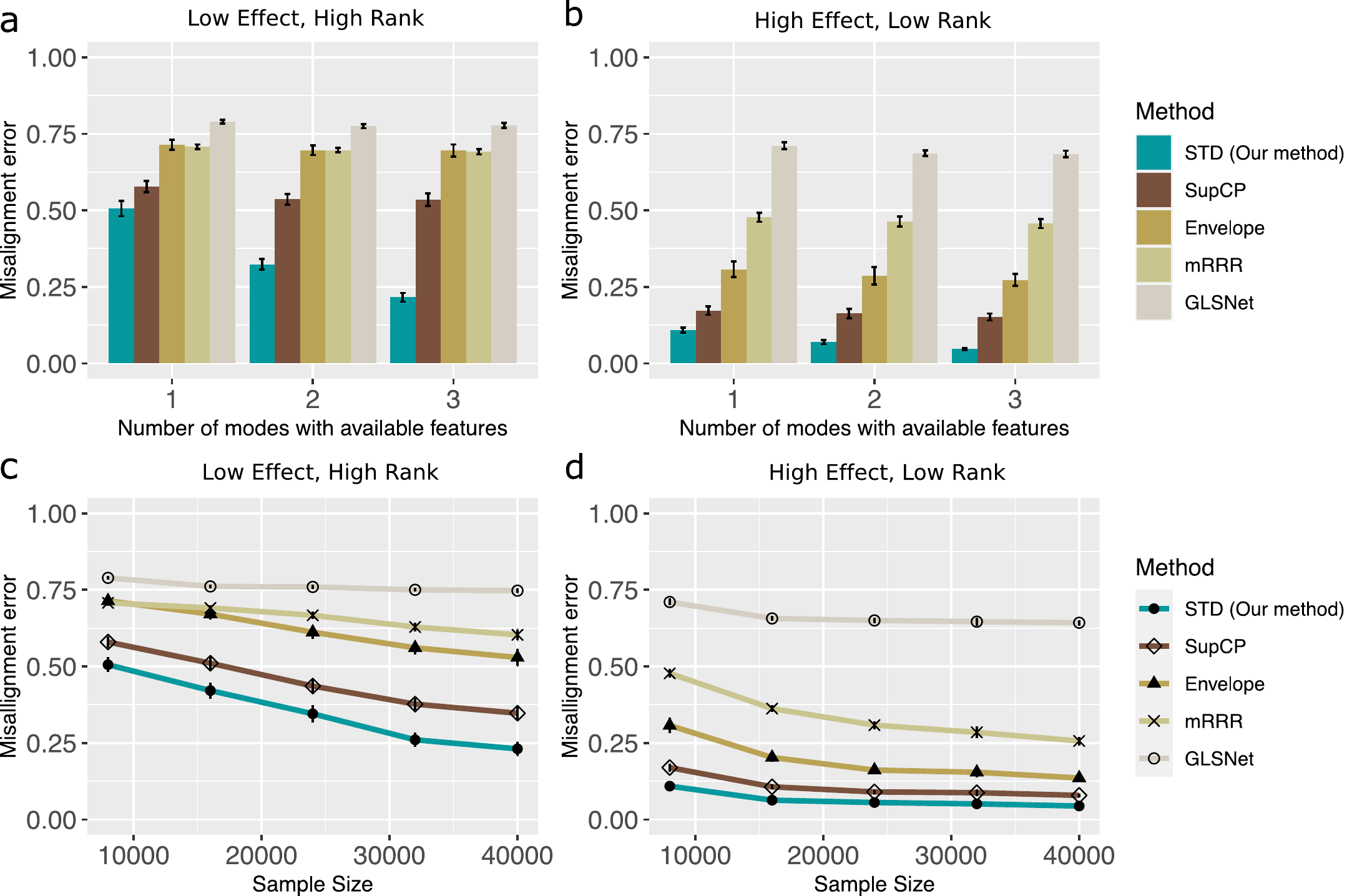} 
\caption{Comparison between tensor methods with Binary data. The panel legends are the same as in Figure~\ref{fig:comp}.}~\label{fig:comp_b}
\vspace{-.5cm}
\end{figure}

We also evaluate the performance comparison with Bernoulli tensors. Figure~\ref{fig:comp_b} indicates the necessarity of generalized model in addressing non-Gaussian data. Indeed, methods that assume Gaussiannity (\textbf{Envelope} and \textbf{SucCP}) perform less favorably in Bernoulli setting (Figure~\ref{fig:comp_b}c) compared to Gaussian setting (Figure~\ref{fig:comp}c). Our method shows improved accuracy as the number of informative features increases (Figure~\ref{fig:comp_b}a-b). In the absence of multiple features, our method still performs favorably compared to others (Figure~\ref{fig:comp_b}c-d), for the same reasons we have argued in Gaussian data. 

\begin{figure}[t]
\centering
\includegraphics[width=12cm]{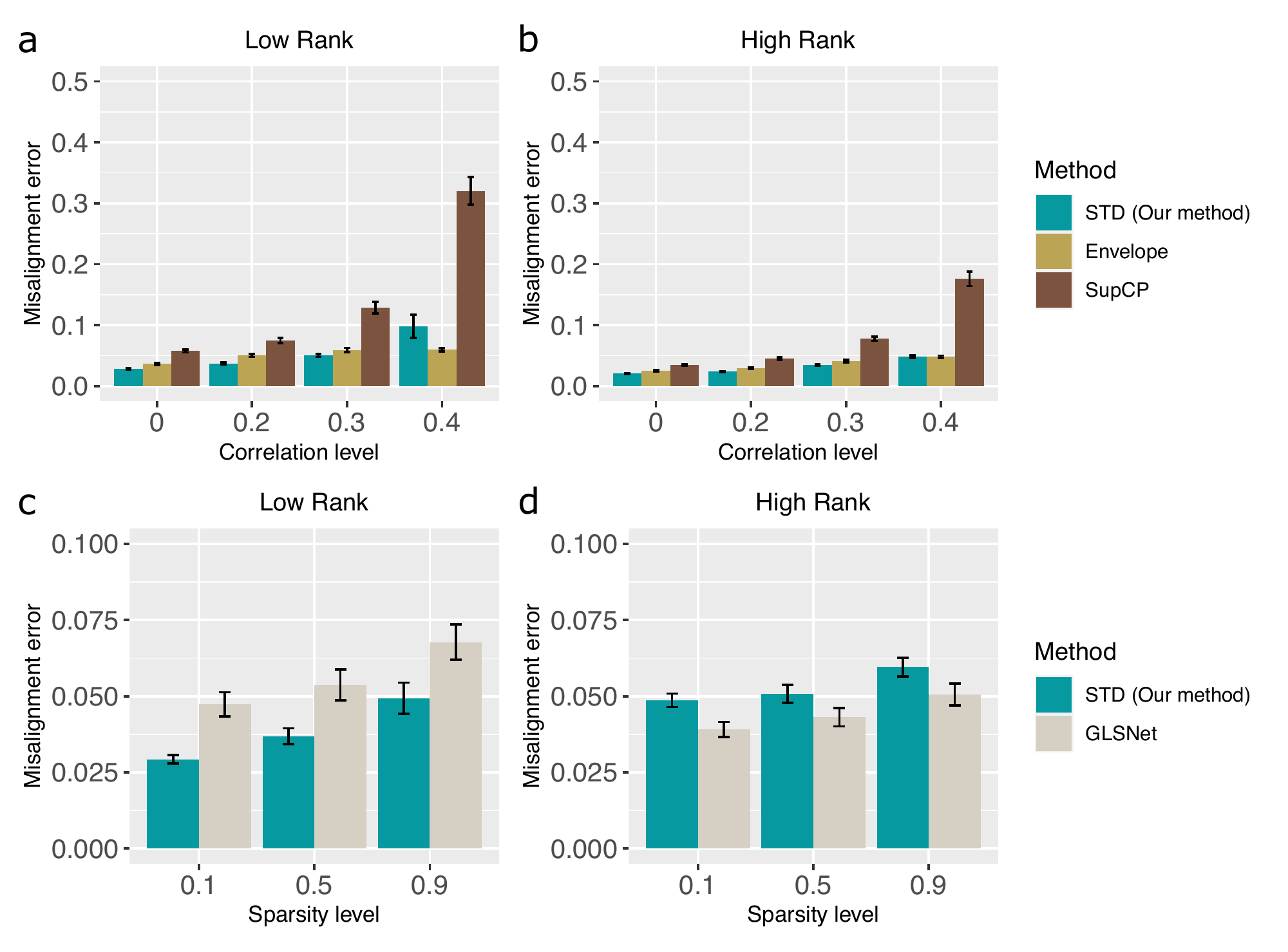} 
\caption{Comparison between tensor methods under model misspecification. Panels (a)-(b) assess the noise correlation, and panels (d)-(d) assess the sparsity. }~\label{fig:noniid}
\vspace{-.5cm}
 \end{figure}

Lastly, we assess the performance of our method {\bf STD} under model misspecification. We consider two aspects: (i) non-independent noise, and (ii) sparse feature effects. Note that our method {\bf STD} imposes neither of these two assumptions, so the experiment allows us to assess the robustness. We select competing methods from Table~\ref{table:comp_table} that specifically addresses these two aspects. We use {\bf Envelope} and {\bf SupCP} as benchmark for noise correlation experiment, and {\bf GLSNet} for sparsity experiment.

Figure~\ref{fig:noniid}a-b assesses the impact of noise correlation to the estimation accuracy. The data is simulated from {\bf Envelope} model with envelope dimensions $r=(3,3)$ (low) and $(4,5)$ (high). The noise is generated from a zero-mean Gaussian tensor with Kronecker structured covariance; see Appendix Section~\ref{sec:asimulation} for details. As expected, {\bf Envelope} shows the best performance in the high correlation setting. Remarkably, we find that our method {\bf STD} has comparable and sometimes better performance when noise correlation is moderate-to-low. In contrast, {\bf SupCP} appears less suitable in this setting. Although {\bf SupCP} allows noise correlation implicitly through latent random factors, the induced correlation may not belong to the Kronecker covariance structure in the simulation.

Figure~\ref{fig:noniid}c-d assesses the impact of sparsity to estimation performance. We generate data from {\bf GLSNet} model, except that we modify the coefficient tensor to be joint sparse and low-rank (the original {\bf GLSNet} model assumes full-rankness on the coefficient tensor). 
The sparsity level ($x$-axis in Figure~\ref{fig:noniid}c-d) quantifies the proportion of zero entries in the coefficient tensor. Since neither our method {\bf STD} nor {\bf GLSNet} follow the simulated model, this setting allows a fair comparison. We find that our method outperforms {\bf GLSNet} in the low-rank setting, whereas {\bf GLSNet} shows a better performance in the high-rank setting. This observation suggests the robustness of our method to sparsity when the tensor of interest is simultaneously low-rank and sparse. When sparsity is the only salient structure, then methods specifically addressing sparsity would provide a better fit. }

\section{Data applications}\label{sec:data}
We apply our supervised tensor decomposition to two datasets. The first application studies the brain networks in response to individual attributes (i.e.,\ feature on one mode), and the second application focuses on multi-relational network analysis with dyadic attributes (i.e.,\ features on two modes). 

\subsection{Application to human brain connection data} 

The Human Connectome Project (HCP) aims to build a network map that characterizes the anatomical and functional connectivity within healthy human brains~\citep{van2013wu}. We follow the preprocessing procedure as in~\citep{desikan2006automated} and parcellate the brain into 68 regions of interest. The dataset consists of 136 brain structural networks, one for each individual. Each brain network is represented as a 68-by-68 binary matrix, where the entries encode the presence or absence of fiber connections between the 68 brain regions. We consider four individual features: gender (65 females vs.\ 71 males), age 22-25 ($n=35$), age 26-30 ($n=58$), and age 31+ ($n=43$). The preprocessed dataset is released in our R package \texttt{tensorregress}. The goal is to identify the connection edges that are affected by individual features. A key challenge in brain network is that the edges are correlated; for example, the nodes in edges may be from a same brain region, and it is of importance to take into account the within-dyad dependence. 

\begin{figure}[!h]
\centering
\includegraphics[width=15cm]{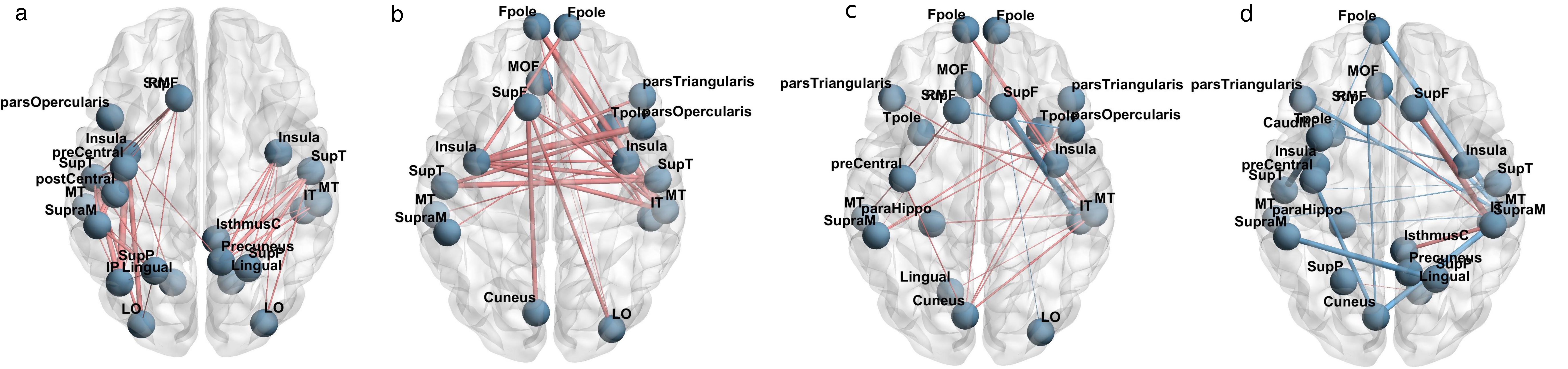}
\caption{Top edges with large effects. (a) Global effect; (b) Female effect; (c) Age 22-25; (d) Age 31+. Red edges represent positive effects and blue edges represent negative effects. The edge-width is proportional to the magnitude of the effect size.
}\label{fig:brain}\label{fig:s1}
\vspace{-.5cm}
\end{figure}

We perform the supervised tensor decomposition to the HCP data. 
The BIC selection suggests a rank $\mr=(10,10,4)$ with quasi log-likelihood $\tL_{\tY}=-174654.7$. We utilize the sum-to-zero contrasts in coding the feature effects, and depict only the top 3\% edges whose connections are non-constant across the sample. Figure~\ref{fig:brain} shows the top edges with high effect size, overlaid on the Desikan atlas brain template~\citep{desikan2006automated}. We find that the global connection exhibits clear spatial separation, and that the nodes within each hemisphere are more densely connected with each other (Figure~\ref{fig:brain}a). In particular, the superior-temproal (\emph{SupT}), middle-temporal (\emph{MT}) and Insula are the top three popular nodes in the network. Interestingly, female brains display higher inter-hemispheric connectivity, especially in the frontal, parental and temporal lobes (Figure~\ref{fig:brain}b). This is in agreement with a recent study showing that female brains are optimized for inter-hemispheric communication~\citep{ingalhalikar2014sex}. We find several edges with declined connection in the group Age 31+. Those edges involve Frontal-pole (\emph{Fploe}), superior-frontal (\emph{SupF}) and Cuneus nodes. The Frontal-pole region is known for its importance in memory and cognition, and the detected decline with age further highlights its biological importance. 

\subsection{Application to political relation data}

The second application studies the multi-relational networks with node-level attributes. We consider \emph{Nations} dataset~\citep{nickel2011three} which records 56 relations among 14 countries between 1950 and 1965. The multi-relational networks can be organized into a $14 \times 14 \times 56$ binary tensor, with each entry indicating the presence or absence of an action, such as ``sending tourist to'', ``export'', ``import'', between countries. The 56 relations span the fields of politics, economics, military, religion, etc. In addition, country-level attributes are also available, and we focus on the following six features: \emph{constitutional, catholics, law ngo, political leadership, geography}, and \emph{medicine ngo}. The goal is to identify the variation in connections due to country-level attributes and their interactions. 

We apply our tensor model to the \emph{Nations} data. The multi-relational network $\tY$ is a binary data tensor, and the country attributes $\mX\in\mathbb{R}^{14\times 6}$ are features on both the 1$^\text{st}$ and 2$^\text{nd}$ modes. {  We use BIC as guidance to select the rank of coefficient tensor $\tB$. Since several rank configurations give similar BIC values, we present here the most interpretable results with $\mr=(4,4,4)$. Detailed rank selection procedure is in Appendix Section~\ref{sec:adata}.}
We perform the supervised tensor decomposition and obtain the dimension reduction matrices $\hat \mM_k$ from the model. Then we apply $K$-mean clustering to dimension reduction matrix on each of the modes. Appendix Table~\ref{tab:s1} shows the $K$-means clustering of the 56 relations based on the dimension reduction matrix on the 3$^\text{rd}$ mode. We find that the relations reflecting the similar aspects of actions are grouped together. In particular, Cluster I consists of military relations such as \emph{violentactions}, \emph{warnings} and \emph{militaryactions}; Clusters II and III capture the economic relations such as \emph{economicaid, booktranslations, tourism}; and Cluster IV represents the political alliance and territory relations. 

\begin{figure}[!h]
\centering
\includegraphics[width=15cm]{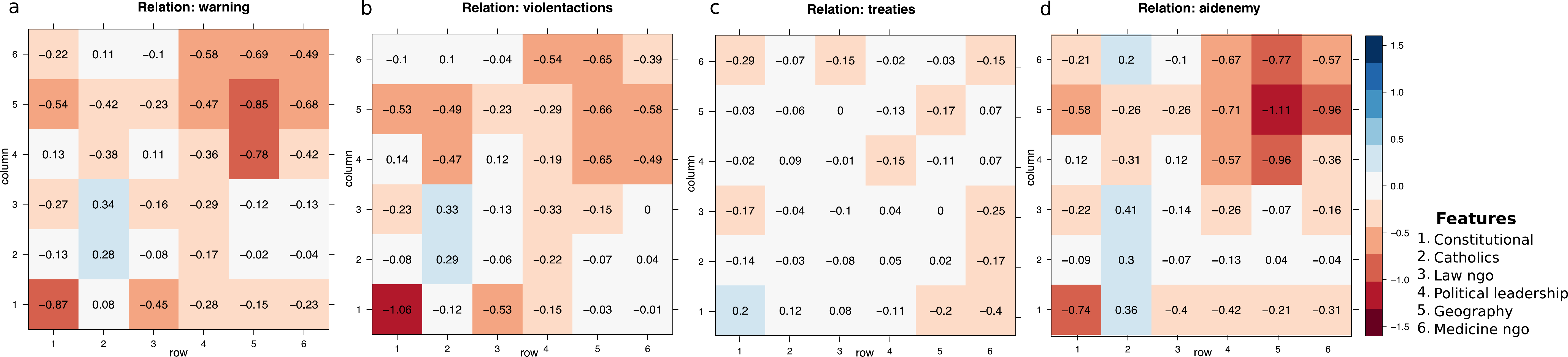}
\caption{Estimated feature effects in the \emph{Nations} data analysis. Panels (a)-(d) represent the estimated effects of country-level attributes towards the connection probability, for relations \emph{warning}, \emph{violentactions}, \emph{treaties}, and \emph{aidenemy}, respectively. }\label{fig:est}
\end{figure}

To investigate the effects of dyadic attributes towards connections, we depict the estimated coefficients $\hat \tB=\entry{\hat b_{ijk}}$ for several relation types (Figure~\ref{fig:est}). { The entry $\hat b_{ijk}$ estimates the contribution, at the logit scale, of feature pair $(i,j)$ ($i$th feature for the ``sender'' country and $j$th feature for the ``receiver'' country) towards the connection of relation $k$.} Several interesting findings emerge from the observation. We find that relations belonging to a same cluster tend to have similar feature effects. For example, the relations ``warning'' and ''violentactions'' are classified into Cluster I, and both exhibit similar effect patterns (Figure~\ref{fig:est}a-b). Moreover, the feature \emph{constitutional} has a strong effect in the relation ``violentactions'' and ``warning'', whereas the effect is weaker in the relation ``treaties''. The result is plausible because the constitutional attributes affect political actions more often than economical actions. The entries in $\tB$ are useful for revealing interaction effects in a context-specific way. 
From Figure~\ref{fig:est}, we find a strong interaction between \emph{geography} and \emph{political leadership} in the relation ``warning'', and a strong interaction between \emph{geogrphy} and \emph{medicine ngo} in the relation ``aidenemy''. The relation-specific effect pattern showcases the applicability of our method in revealing complex interactions. 

\vspace{-0.5cm}

\section{Proofs}\label{sec:proof}
Here, we provide the proofs for Theorem~\ref{thm:MLE}, Proposition~\ref{lem:ini} and Theorem~\ref{thm:local}, and auxiliary lemmas. 

\subsection{Proof of Theorem~\ref{thm:MLE}}\label{subsec:proof_MLE}
We denote several quantities:
\begin{equation}\label{eq:quantities}
\underline \gamma = \prod_{k\in[K]} \sigma_{\min}(\mX_k),\quad \bar \gamma =\prod_{k\in[K]}\sigma_{\max}(\mX_k),  \quad \lambda =\min_{k\in[K]}\sigma_{\min}(\text{Unfold}_k(\tB_{\text{true}})),
\end{equation}
where $\underline{\gamma}$ quantifies the rank non-deficiency of feature matrices, $\bar \gamma$ quantifies the magnitude of feature matrices, and $\lambda$ is the smallest singular value of mode-$k$ unfolded matrices $\text{Unfold}_k(\trueB)$ for all possible $k\in[K]$. 
For notational convenience, we drop the subscript $\tY$ from the objective $\tL_\tY(\cdot)$ and simply write as $\tL(\cdot)$. We write $\tL(\tB)$ in place of $\tL(\tC, \mM_1,\ldots,\mM_K)$ when we want to emphasize the role of $\tB$. 

\begin{prop}[sub-Gaussian residuals]\label{prop}
Define the residual tensor $\tE=\entry{\varepsilon_{i_1,\ldots,i_K}}=\tY-b'(\Theta)\in\mathbb{R}^{d_1\times \cdots \times d_K}$. Under the Assumption A2, $\varepsilon_{i_1,\ldots,i_K}$ is a sub-Gaussian random variable with sub-Gaussian parameter bounded by $\phi U$, for all $(i_1,\ldots,i_K)\in[d_1]\times\cdots\times[d_K]$.
\end{prop}

\begin{prop}[Properties of tensor GLM]\label{lem}Consider tensor GLMs under Assumption A2.
\begin{enumerate}[label=(\alph*)]
\item (Strong convexity) For all $\tB$ and all realized data tensor $\tY$,
\[
\tL(\trueB)\geq \tL(\tB)+ \langle \nabla \tL(\tB_{\text{true}}), \trueB-\tB \rangle+{1\over 2} \underline{\gamma}^2L\FnormSize{}{\trueB-\tB}^2,
\] 
where $\nabla L(\cdot)$ denotes the derivative of $\tL$ with respect to $\tB$.   
\item (Model complexity) 
Suppose $\tY$ follows generalized tensor model with parameter $\trueB$. Then, with probability at least $1-\exp(-p)$, 
\begin{equation}\label{eq:b}
\text{Err}_{\text{ideal}}(\mr):= \sup_{\substack{ \FnormSize{}{\tB}=1, \tB\in\tP(\mr)}}\langle \nabla \tL(\tB_{\text{true}}), \tB \rangle\lesssim \bar \gamma \sqrt{\phi U(r^K+Kpr)}. 
\end{equation}
\end{enumerate}
\end{prop}
The proofs of Propositions~\ref{prop}-\ref{lem} are in Section~\ref{sec:lemma}. 

\begin{proof}[Proof of Theorem~\ref{thm:MLE}]
First we prove the error bound for $\MLEB$. By the definition of $\MLEB$, $\tL_{\tY}(\trueB)-\tL_{\tY}(\MLEB)\leq 0$. By the strong convexity in Proposition~\ref{lem},
\begin{align}\label{eq:F-norm}
0&\geq \tL_{\tY}(\trueB)- \tL_{\tY}(\MLEB)  \geq \langle \nabla \tL(\tB_{\text{true}}), \trueB-\MLEB \rangle+{1\over 2}\underline{\gamma}^2L\FnormSize{}{\trueB-\MLEB}^2.
\end{align}
Rearranging~\eqref{eq:F-norm} gives
\[
\FnormSize{}{\MLEB-\trueB}\leq {2\over \underline{\gamma}^2L} \left\langle \nabla \tL(\tB_{\text{true}}), {\MLEB-\trueB\over\FnormSize{}{\MLEB-\trueB}} \right\rangle\leq {2\over \underline{\gamma}^2L}  \text{Err}_{\text{ideal}}(2\mr),
\]
where the last inequality comes from the definition of $\text{Err}_{\text{ideal}}(2\mr)$ and the fact that $\text{rank}(\MLEB-\trueB)\leq \text{rank}(\MLEB)+\text{rank}(\trueB)\leq 2\mr$. By~\eqref{eq:b} in Proposition~\ref{lem}, we have 
\begin{equation}\label{eq:Bbound}
\FnormSize{}{\MLEB-\trueB} \lesssim {\bar \gamma \sqrt{\phi U}\over \underline{\gamma}^2 L}\sqrt{r^K+Kpr},
\end{equation}
with probability at least $1-\exp(-p)$.

Now, we specialize $\bar \gamma/\underline{\gamma}^2 $ in the following two cases of assumptions on feature matrices. 
\begin{itemize}[leftmargin=0cm]
\item[] [Case 1] Under Assumption A1 with scaled feature matrices, we have
\begin{equation}\label{eq:case1}
{\bar \gamma\over \underline{\gamma}^2} \leq {c^K_2d^{K/2} \over c^{2K}_1d^{K}} \lesssim \sqrt{1\over d^{K}}.
\end{equation}
\item[] [Case 2] Under Assumption A1' with original feature matrices, the asymptotic behavior of extreme singular values \citep{rudelson2010non} are
\[
\sigma_{\min}(\mX_k) \asymp \sqrt{d}-\sqrt{p} \text{ and  } \sigma_{\max}(\mX_k) \asymp \sqrt{d}+\sqrt{p},\quad \text{for all }k\in[K].
\]
In this case, we obtain 
\begin{equation}\label{eq:case2}
{\bar \gamma\over \underline{\gamma}^2} \asymp {(\sqrt{d}+\sqrt{p})^{K} \over (\sqrt{d}-\sqrt{p})^{2K}} \lesssim \sqrt{1\over d^K}.
\end{equation}
\end{itemize}
Combining~\eqref{eq:Bbound} with either~\eqref{eq:case1} or~\eqref{eq:case2}, in both cases we obtain the same conclusion
\begin{equation}\label{eq:bound2}
\FnormSize{}{\MLEB-\trueB}^2\lesssim {\phi(r^K+Kpr)\over d^K}.
\end{equation}
Now we prove the bound for sin$\Theta$ distance. 
We unfold tensors $\trueB$ and $\MLEB$ along the mode $k$ and obtain $\text{Unfold}_k(\trueB)$ and $\text{Unfold}_k(\MLEB)$. Notice that $\trueM$ and $\hat \mM_{k,\textup{MLE}}$ span the top-$r$ left singular spaces of $\text{Unfold}_k(\trueB)$ and $\text{Unfold}_k(\MLEB)$, respectively.  Applying Proposition~\ref{prop:sinebound} to this setting gives
\begin{align}\label{eq:sintheta}
    \text{sin}\Theta(\trueM,\hat\mM_{k,\textup{MLE}})\leq \frac{\FnormSize{}{\text{Unfold}_k(\MLEB)-\text{Unfold}_k(\trueB)}}{\sigma_{\text{min}}(\text{Unfold}_k(\trueB))}= \frac{\FnormSize{}{\MLEB-\trueB}}{\lambda}.
    \end{align}
The proof is complete by combining~\eqref{eq:bound2} and~\eqref{eq:sintheta}.
\end{proof}

\subsection{Proofs of Proposition~\ref{lem:ini} and Theorem~\ref{thm:local}}\label{subsec:proof_ini}
\begin{proof}[Proof of Proposition~\ref{lem:ini}]

We express the Gaussian model as 
\begin{align}\label{eq:gaussian}
    \tY = \tB_{\text{true}}\times \{\mX_1,\ldots,\mX_K\} + \tE,
\end{align}
where $\tE$ is a noise tensor consisting of i.i.d.\ entries from $N(0,\sqrt{\phi}).$ 
By QR decomposition on feature matrices, $\mX_k = \mQ_k\mR_k$ for all $k\in[K]$, we have 
\begin{align}\label{eq:gaussian2}
    \bar\tY = \tB_{\text{true}}\times\{\mR_1,\ldots \mR_K\}+\bar\tE,
\end{align}
where $\bar\tY = \tY\times\{\mQ_1,\ldots,\mQ_K\}$ and  $\bar\tE = \tE\times\{\mQ_1,\ldots,\mQ_K\}$.
Notice that entries of $\bar\tE\in\mathbb{R}^{p\times\cdots\times p}$ are i.i.d drawn from $N(0,\sqrt{\phi})$ by the orthonormality of $\{\mQ_k\}_{k=1}^K$. Reparameterize the signal in \eqref{eq:gaussian2} as 
\begin{align}\label{eq:repara}
    \tS_{\text{true}} := \tB_{\text{true}}\times\{\mR_1,\ldots \mR_K\}&=  \tC_{\text{true}}\times\{\mR_1\mM_{1,\text{true}},\ldots \mR_K\mM_{K,\text{true}}\}\notag \\
    & = \tC'_{\text{true}}\times\{\mU_{1,\text{true}},
    \ldots,\mU_{K,\text{true}}\},
\end{align}
where $\mU_{k.\text{true}}\in\mathbb{O}(p_k, r_k)$ are orthonormal matrices and $\tC'_{\text{true}}\in\mathbb{R}^{r\times \cdots\times r}$ is a full rank core tensor.
By definition of quantities in \eqref{eq:quantities}, we have
\begin{align}\label{eq:lambdas}
    \lambda' := \min_{k\in[K]}\sigma_{\min}(\text{Unfold}_k(\tS_{\text{true}}))\in[  \lambda \underline \gamma,\ \lambda \bar \gamma].
\end{align}

Now our setup shares the same setting as in~\citet[Theorem 1]{zhang2018tensor}.
We summarize the relationships between our algorithm outputs and the ones in \cite{zhang2018tensor}. For all $k\in[K],$
\begin{enumerate}
    \item  $\mM_{k,\text{true}} = \text{SVD}_{r_k}\left(\mR_k^{-1}\mU_{k,\text{true}}\right):=$  the first $r_k$ left singular vectors of $\mR_k^{-1}\mU_{k,\text{true}}$;
    \item $\hat\mM_{k}^{(t)} = \text{SVD}_{r_k}\left(\mR_k^{-1}\hat\mU_{k}^{(t)}\right)$ for all $t=0,1,2,\ldots$;
\end{enumerate}
where $\hat \mU^{(t)}_k$ denotes the $t$-th iteration output of Higher Order Orthogonal Iteration (HOOI) algorithm \citep{zhang2018tensor} with inputs $\bar \tY$. 
The first relationship is from~\eqref{eq:repara}, and second relationship is from induction by $t$. Briefly, $t=0$ holds because of the definition  $\hat\mM^{(0)}_k$ based on lines 4-5 of our initialization Algorithm~\ref{alg:A}. For $t\geq 1,\ldots$, notice that $\hat\mM_k^{(t)}$ is an optimizer of the objective \[\FnormSize{}{\bar\tY-\hat\tC^{(t-1)}\times\{\mR_1\hat\mM_1^{(t)},\ldots,\mR_{k-1}\hat\mM_{k-1}^{(t)},\mR_k\mM,\mR_{k+1}\hat\mM_{k+1}^{(t-1)},\ldots,\mR_K\hat\mM^{(t-1)}_K\}}^2,\]
from the line 3 of  Algorithm~\ref{alg:B}. By unfolding along the mode $k$, the optimizer $\mM_k^{(t)}$ must satisfy
\begin{align}\label{eq:HOOI}
    \text{Unfold}_k&\left(\bar\tY\times\left\{ (\hat\mM_1^{(t)})^T\mR_1^{-1},\ldots, (\hat\mM_{k-1}^{(t)})^T\mR_{k-1}^{-1},\ \mI_{p_k}, \ (\hat\mM_{k+1}^{(t-1)})^T\mR_{k+1}^{-1}, \ldots, (\hat\mM_{K}^{(t-1)})^T\mR_{K}^{-1}\right\} \right)\notag \\
    &=\mR_k\hat\mM_k^{(t)}\text{Unfold}_k\left(\hat\tC^{(t-1)}\right)\left(\mI_{r_K}\otimes\cdots\otimes\mI_{r_{k+1}}\otimes\mI_{r_{k-1}}\otimes \mI_{r_1}\right).
\end{align}
Notice that the first $r_k$ left singular vectors of the left side of~\eqref{eq:HOOI} is $\hat\mU^{(t)}_k$ in HOOI algorithm. Therefore, we prove the  the second relationship by induction.

Combination of Lemma~\ref{lem:sint} and the relationships between our algorithm outputs and the ones in \cite{zhang2018tensor} gives us
\begin{align}\label{eq:sinrel}
    \left(\frac{\underline\gamma}{\bar\gamma}\right)^2\max_{k\in[K]}\sin\Theta(\mU_{k,\text{true}},\hat\mU_k^{(t)})\leq
    \max_{k\in[K]}\sin\Theta(\mM_{k,\text{true}},\hat\mM_k^{(t)})\leq \left(\frac{\bar\gamma}{\underline\gamma}\right)^2\max_{k\in[K]}\sin\Theta(\mU_{k,\text{true}},\hat\mU_k^{(t)}).
\end{align}
Now, we prove the property (a) in Proposition~\ref{lem:ini}. Based on Lemma~\ref{lem:lts}(a), whenever $\lambda'/\sqrt{\phi}\geq C_{\text{gap}}p^{K/4}$, we have
\begin{align}\label{eq:inim}
    \max_{k\in[K]}\sin \Theta(\mU_{k,\text{true}},\hat\mU_k^{(0)}) \leq c\left(\frac{p^{K/2}}{(\lambda\underline\gamma)^2/\phi}\right),
\end{align}
 with probability at least $1-\exp(-p)$.
Notice that
\begin{align}\label{eq:cgap}
    \lambda'\stackrel{\eqref{eq:lambdas}}{\geq} \lambda \underline \gamma\gtrsim \lambda d^{K/2}\geq C_{\text{gap}}\sqrt{\phi} p^{K/4},
\end{align}
where the second inequality uses [Case 1] and [Case 2] in the proof of Theorem~\ref{thm:MLE}. 
The condition $\lambda/\sqrt{\phi}\geq C p^{K/4} d^{-K/2}$ guarantees  a sufficiently large $C_{\text{gap}}$ that satisfies $\lambda'/\sqrt{\phi}\geq C_{\text{gap}}p^{K/4}$.
Thus combining \eqref{eq:sinrel} and \eqref{eq:inim} yields
\begin{align}
    \max_{k\in[K]}\sin\Theta(\mM_{k,\text{true}},\hat\mM_k^{(0)}) &
    \leq \left(\frac{\bar\gamma}{\underline\gamma}\right)^2\left(\sqrt{\phi}p^{K/4}\over\lambda\underline\gamma\right)^2
    \\
    &\leq \frac{1}{2},
\end{align}
where the last inequality uses the fact that  $\underline\gamma\asymp d^{K/2}$ and $\bar\gamma/\underline\gamma$ is bounded by a constant in [Case 1] and [Case 2], and the condition $\lambda /\sqrt{\phi}\geq Cp^{K/4}d^{-K/2}$.

Now, we prove the property (b) in Proposition~\ref{lem:ini}. Based on Lemma~\ref{lem:lts}(b), we have
 \begin{align}\label{eq:lt2}
      \max_{k\in[K]}\sin\Theta(\mU_{k,\text{true}},\hat\mU_k^{(t)})\lesssim \frac{\sqrt{p\phi}}{\lambda\underline\gamma}+ \left(1\over 2\right)^t\max_{k\in[K]}\sin\Theta(\mU_{k,\text{true}},\hat\mU_k^{(0)}),
 \end{align}
  with probability at least $1-\exp(-p)$.
  Combining \eqref{eq:sinrel} with the above inequality yields
 \begin{align}
     \max_{k\in[K]}\sin\Theta(\mM_{k,\text{true}},\hat\mM_k^{(t)}) 
    &\lesssim \max_{k\in[K]}\sin\Theta(\mU_{k,\text{true}},\hat\mU_k^{(t)})\\&\lesssim \frac{\sqrt{p\phi}}{\lambda\underline\gamma}+ \left(1\over 2\right)^t\max_{k\in[K]}\sin\Theta(\mU_{k,\text{true}},\hat\mU_k^{(0)})\\&\lesssim\frac{\sqrt{p\phi}}{\lambda\underline\gamma}+ \left(1\over2\right)^t\max_{k\in[K]}\sin\Theta(\mM_{k,\text{true}},\hat\mM_k^{(0)}).
 \end{align}
Finally, the proof is completed applying $\underline\gamma\asymp d^{K/2}$ from [Case 1] and [Case 2].
\end{proof}

\begin{proof}[Proof of Theorem~\ref{thm:local}]
Combining Proposition~\ref{lem:ini}(b) and \eqref{eq:inim}, we obtain
 \begin{align}
      \max_{k\in[K]}\sin\Theta(\mU_{k,\text{true}},\hat\mU_k^{(t)})\lesssim \frac{\sqrt{p\phi}}{\lambda\underline\gamma}+ \left(1\over 2\right)^t\left(\frac{p^{K/2}}{(\lambda\underline\gamma)^2/\phi}\right),
 \end{align}
 with probability at least $1-\exp(-p)$.
We set $t\gtrsim \log \frac{p^{(K-1)/2}}{\lambda\underline\gamma}$ to make the second term negligible.
 Therefore, the first part of proof is completed by noticing that \[ \frac{p^{(K-1)/2}}{\lambda\underline\gamma}\lesssim\log \frac{p^{(K-1)/2}}{\lambda d^{K/2}}\lesssim K\log  p,\] where the first inequality uses $\underline\gamma\asymp d^{K/2}$ from [Case 1] and [Case 2], and the last inequality is from the condition $\lambda/\sqrt{\phi}\geq Cp^{K/4} d^{-K/2}$. 
 
 For the estimation error with respect to Frobenius norm, direct application of Lemma~\ref{lem:lts}(c) with $t\gtrsim K\log p\gtrsim \log\frac{p^{(K-1)/2}}{\lambda\underline\gamma}$ yields
 \begin{align}\label{eq:sbound}
     \FnormSize{}{\hat\tS^{(t)}-\tS_{\text{true}}}^2\lesssim \phi(r^K+Kpr),
 \end{align}
 with probability at least $1-\exp(-p).$
Notice that
\begin{align}\label{eq:sbbound}
    \FnormSize{}{\hat\tS^{(t)}-\tS_{\text{true}}}^2 &= \FnormSize{}{\left(\hat\tB^{(t)}-\tB_{\text{true}}\right)\times\{\mR_1,\ldots \mR_K\}}^2\nonumber\\&\geq{\underline{\gamma}}^2 \FnormSize{}{\hat\tB^{(t)}-\tB_{\text{true}}}^2\nonumber\\&\gtrsim d^{K}\FnormSize{}{\hat\tB^{(t)}-\tB_{\text{true}}}^2, \quad \text{from [Case 1] and [Case 2]}.
\end{align}
Combining \eqref{eq:sbound} and \eqref{eq:sbbound} completes the proof.
\end{proof}

\subsection{Auxiliary Lemmas}\label{sec:lemma}
\begin{proof}[Proof of Proposition~\ref{prop}]For ease of presentation, we drop the subscript $(i_1,\ldots,i_K)$ and simply write $\varepsilon$ ($=y-b'(\theta)$). For any given $t\in\mathbb{R}$, we have
\begin{align}
\mathbb{E}(\exp(t\varepsilon|\theta))&=\int c(x) \exp\left({\theta x - b(\theta)\over \phi}   \right)\exp \left(t(x-b'(\theta))\right)dx\\
&=\int c(x)\exp \left( {(\theta + \phi t)x - b (\theta+\phi t)+b(\theta+\phi t)-b(\theta)-\phi t b'(\theta) \over \phi}\right)dx\\
&=\exp\left( {b(\theta+\phi t)-b(\theta)-\phi t b'(\theta) \over \phi} \right)\\
&\leq \exp\left(\phi U t^2\over 2 \right),
\end{align}
where $c(\cdot)$ and $b(\cdot)$ are known functions in the exponential family corresponding to $y$, and the last line uses the fact that $\sup_{\theta\in\mathbb{R}}b''(\theta)\leq U$. 
Therefore, $\varepsilon$ is sub-Gaussian-$(\phi U)$. \end{proof}

\begin{defn}[$\alpha$-convexity] A real-valued function $f\colon \tS \to\mathbb{R}$ is called $\alpha$-convex, if
\[
f(x_1)\geq f(x_2)+\langle \nabla_xf(x_2),\ x_1-x_2\rangle+{\alpha}\FnormSize{}{x_1-x_2}^2,\ \text{for all }x_1,x_2\in\tS.
\]
\end{defn}
\begin{lem}[Convexity under linear transformation]\label{lem:convex}
Suppose $f\colon \mathbb{R}^{d\times \cdots\times d}\to \mathbb{R}$ is a $\alpha$-convex function. Define a function $g\colon \mathbb{R}^{p\times \cdots \times p}\to \mathbb{R}$ by
$g(\tB)=f(\tB\times \{\mX_1,\ldots,\mX_K\})$ for all $\tB\in\mathbb{R}^{p\times \cdots\times p}$. Then, $g$ is a $(\underline{\gamma}^2\alpha)$-convex function. 
\end{lem}
\begin{proof}[Proof of Lemma~\ref{lem:convex}]
By the definition of $\alpha$-convexity, we have
\begin{equation}\label{eq:f}
f(\Theta_1)\geq f(\Theta_2) + \langle \nabla_\Theta f(\Theta_2), \Theta_1-\Theta_2\rangle + \alpha\FnormSize{}{\Theta_1-\Theta_2}^2, \ \text{for all }\Theta_1,\Theta_2\in\mathbb{R}^{d\times \cdots\times d},
\end{equation}
where $\nabla_\Theta f(\cdot)$ denotes the derivative of $f$ with respect to $\Theta\in\mathbb{R}^{d\times \cdots\times d}$. For any $\tB_1, \tB_2\in\mathbb{R}^{p\times \cdots \times p}$, we notice that $\tB_i\times\{\mX_1,\ldots,\mX_K\}\in\mathbb{R}^{d\times \cdots \times d}$ for $i=1,2$. Applying~\eqref{eq:f} to this setting gives
\begin{align}\label{eq:1}
&f(\tB_1\times\{\mX_1,\ldots,\mX_K\}) \notag \\
\geq &\ f(\tB_2\times\{\mX_1,\ldots,\mX_K\})+\langle  \nabla_\Theta f(\tB_2\times\{\mX_1,\ldots,\mX_K\}), (\tB_1-\tB_2)\times\{\mX_1,\ldots,\mX_K\} \rangle \notag \\
&\ +  \alpha\FnormSize{}{(\tB_1-\tB_2)\times\{\mX_1,\ldots,\mX_K\}}^2\notag \\
\geq &\ f(\tB_2\times\{\mX_1,\ldots,\mX_K\})+\langle  \nabla_\Theta f(\tB_2\times\{\mX_1,\ldots,\mX_K\})\times\{\mX^T_1,\ldots,\mX^T_K\}, (\tB_1-\tB_2)\rangle \notag \\
&\ + \alpha \underline{\gamma}^2  \FnormSize{}{\tB_1-\tB_2}^2.
\end{align}
By the definition of $g$ and the linearity from $\tB$ to $\Theta$, we have
\begin{equation}\label{eq:2}
\nabla g_\tB(\tB_2)=\nabla f_\Theta (\tB_2\times\{\mX_1,\ldots,\mX_K\}) \times\{\mX^T_1,\ldots,\mX^T_K\}.
\end{equation}
The convexity of $g$ directly follows by plugging~\eqref{eq:2} into~\eqref{eq:1},
\[
g(\tB_1)\geq g(\tB_2)  +\langle \nabla g_\tB(\tB_2), \tB_1-\tB_2\rangle + \alpha \underline{\gamma}^2  \FnormSize{}{\tB_1-\tB_2}^2.
\]
\end{proof}

\begin{proof}[Proof of Proposition~\ref{lem}] We first prove the strong concavity by viewing the log-likelihood as a function of the linear predictor $\Theta$. Write
\[
\bar \tL(\Theta)=\langle \tY, \Theta \rangle -\sum_{i_1,\ldots,i_K}b(\theta_{i_1,\ldots,i_K}).
\]
Direct calculation shows that the Hession of $\bar \tL(\Theta)$ can be expressed as
\[
{\partial \bar \tL(\Theta) \over \partial \theta_{i_1,\ldots,i_K}\partial \theta_{j_1,\ldots,j_K}}=
\begin{cases}
 -b''(\theta_{i_1,\ldots,i_K})<-L<0,\quad \text{if } (i_1,\ldots,i_K)= (j_1,\ldots,j_K),\\
 0, \quad \text{otherwise},
 \end{cases}
\]
Therefore, the Hession matrix of $\bar \tL(\Theta)$ is strictly negative definite with eigenvalues upper bounded by $-L<0$. By Taylor expansion, $-\bar \tL(\Theta)$ is $L/2$-convex with respect to $\Theta$. Note that $\bar \tL(\Theta)=\tL(\tB)$ via the linear mapping $\Theta=\tB\times\{\mX_1,\ldots,\mX_K\}$. Therefore, by Lemma~\ref{lem:convex}, $\tL(\tB)$ is $(\gamma^2L/2)$-convex with respect to $\tB$. 

To prove the second part of Proposition~\ref{lem}, we note
\[
\langle \nabla \tL(\trueB), \tB\rangle=\langle \nabla \bar\tL(\trueT)\times\{\mX^T_1,\ldots,\mX^T_K\},\ \tB \rangle =\langle \tY-b'(\trueT),\ \tB\times\{\mX_1,\dots,\mX_K\}\rangle.
 \]
By Proposition~\ref{prop}, $\tY-b'(\trueT)$ is a random tensor consisting of i.i.d.\ sub-Gaussian-($U\phi$) entries under Assumption 2. We write $\tE=\tY-b'(\trueT)$ and consider the sub-Gaussian maxima
\[
\text{Err}_{\textup{ideal}}(\mr)=\sup_{\FnormSize{}{\tB}=1, \tB\in\tP(r)}\langle \tE, \tB\times\{\mX_1,\ldots,\mX_K\}\rangle.
\]
The quantity $\text{Err}_{\textup{ideal}}(\mr)$ is closely related to the localized Gaussian width~\citep{chen2019non,han2020optimal} that measures the model complexity of $\tP(\mr)$. By adapting \citet[Lemma E.5]{han2020optimal} in our context, we have
\[
\text{Err}_{\textup{ideal}}(\mr) \lesssim  \sqrt{\phi U(r^K+Kpr)}\prod_{k\in[K]} \sigma_{\max}({\mX_k}) \leq \bar \gamma  \sqrt{\phi U(r^K+Kpr)},
\]
with probability at least $1-\exp(-p)$. 
\end{proof}

The following Lemma is adopted from~\citet[Theorem 6.1]{wang2017tensor} in our contexts. 
\begin{lem}[Wedin's $\sin\Theta$ Theorem]\label{prop:sinebound}
Let $\mB$ and $\hat\mB$ be two $m\times n$ real matrix SVDs $\mB = \mU\Sigma\mV^T$ and $\hat\mB = \hat\mU\hat\mSigma\hat\mV^T.$ If $\sigma_{\text{min}}(\mB)>0$ and $\FnormSize{}{\hat\mB-\mB}\ll \sigma_{\text{min}}(\mB)$, then
\begin{align}\label{eq:sine}
    \text{sin} \Theta(\mU,\hat\mU) \leq \frac{\sigma_{\max}(\hat\mB-\mB)}{\sigma_{\text{min}}(\mB)}\leq \frac{\FnormSize{}{\hat\mB-\mB}}{\sigma_{\text{min}}(\mB)}.
\end{align}
\end{lem}

The following theorem~\cite{zhang2018tensor} provides the statistical guarantees for unsupervised tensor decomposition based on alternating least square algorithm. For simplicity, we consider the balanced dimension $p_1 =\cdots=p_K =p$ and $r_1 =\cdots=r_K =r$. 
\begin{lem}[Theorem 1 in \cite{zhang2018tensor}]\label{lem:lts}
Consider the Gaussian tensor model
\begin{align}
    \tY  = \tS_{\text{true}} + \tE,
\end{align}
where $\tS_{\text{true}} = \tC_{\text{true}}\times\{\mU_{1,\text{true}},\ldots,\mU_{K,\text{true}}\}$ is an unknown signal tensor, $\tC_{\text{true}}\in\mathbb{R}^{r\times\cdots\times r}$ is a full rank core tensor, $\mU_{k,\text{true}}\in\mathbb{O}(p,r)$ are orthornomal matrices, and $\tE\in\mathbb{R}^{p\times \cdots \times p}$ is a Gaussian noise tensor consisting of i.i.d entries from $N(0,\sigma)$. Let $\lambda$ denote the smallest singular value of matrices $\text{Unfold}_k(\tS_{\text{true}})$ over all possible $k$, 
\[
\lambda' =\min_{k\in[K]}\sigma_{\min}(\text{Unfold}_k(\tS_{\text{true}})).
\]
Then, the following two properties hold whenever $\lambda'/\sigma\geq C_{\textup{gap}}p^{K/4}$ for some universal constant $C_{\textup{gap}}>0$. 

\begin{enumerate}[label=(\alph*)]
\item  With probability at least $1- \exp(-p)$, the spectral initialization $\hat\mU_k^{(0)}$ has
\begin{align}
   \max_{k\in[K]}\sin \Theta(\mU_{k,\text{true}},\hat\mU_k^{(0)})\leq c \frac{p^{K/2}}{\lambda'^2/\sigma^2},
 \end{align} 
 for some constant $c>0.$

  \item Let $t = 1,2, \ldots,$ denote the  iteration in HOOI algorithm.  With probability at least $1- \exp(-p)$, the alternating optimization $\hat\mU_k^{(t)}$ satisfies
 \begin{align}\label{eq:lt}
      \max_{k\in[K]}\sin \Theta(\mU_{k,\text{true}},\hat\mU_k^{(t)})\lesssim \frac{\sqrt{p}}{\lambda'/\sigma}+ \left({1\over 2}\right)^t\max_{k\in[K]}\sin \Theta(\mU_{k,\text{true}},\hat\mU_k^{(0)}),
 \end{align}
 \item When $t\gtrsim \log {p^{(K-1)/2}\over \lambda'}$, the tensor estimate $\hat \tS^{(t)}$ from HOOI satisfies
 \[
\FnormSize{}{\hat \tS^{(t)}-\tS_{\text{true}}}^2\lesssim  \sigma^2 (r^K+Kpr),
 \]
 with probability at least $1-\exp(-p)$. 
 \end{enumerate}
\end{lem}

\begin{lem}[Angle distance under linear transformation]\label{lem:sint}
Let $\mU$ and $\hat\mU$ be two $m\times n$ real matrices where $m>n$.  Let $\mR$ be an $m\times m$ invertible matrix. If $\sin\Theta(\mU,\hat\mU)\leq L$ for some constant $L\in[0,1]$, then
\begin{align}
    \sin \Theta(\mR\mU,\mR\hat\mU)\leq\left(\sigma_{\max}(\mR)\over \sigma_{\min}(\mR)\right)^2 L.
\end{align}
\end{lem}
\begin{proof}
Suppose that orthonormal basis of $\text{Span}(\mU)$  and $\text{Span}(\hat\mU^{\perp})$ are $\{\mu_1,\ldots,\mu_n\}$ and $\{\nu_{n+1},\ldots,\nu_{m}\}$ respectively. By definition, 
\begin{align}
    \sin \Theta(\mU,\hat\mU) = \max_{\sum_{i=1}^n a_i^2 = \sum_{j=n+1}^m b_j^2 = 1} \left\langle \sum_{i=1}^n a_i\mu_i,\sum_{j=n+1}^m b_j\nu_j\right\rangle \leq L.
\end{align}
We write $\mx = \mR\sum_{i=1}^n a_i\mu_i$ and $\my = \mR\sum_{j=n+1}^m b_j\nu_j$ for any $\mx\in \text{Span}(\mR\mU)$ and $\my\in\text{Span}((\mR\hat\mU)^{\perp})$.
Then,
\begin{align}
    {\langle \mx,\my\rangle\over \|\mx\|_2\|\my\|_2} &= {\langle  \mR\sum_{i=1}^n a_i\mu_i, \mR\sum_{j=n+1}^m b_j\nu_j\rangle\over \|\mR\sum_{i=1}^n a_i\mu_i\|_2\|\mR\sum_{j=n+1}^m b_j\nu_j\|_2 }\\&\leq {\sigma_{\max}(\mR^T\mR)\langle \sum_{i=1}^n a_i\mu_i,\sum_{j=n+1}^m b_j\nu_j\rangle\over \sigma_{\min}^2(\mR)\sqrt{ \sum_{i=1}^n a_i^2}\sqrt{\sum_{j=n+1}^m b_j^2} }\\&\leq \left(\sigma_{\max}(\mR)\over \sigma_{\min}(\mR)\right)^2 \sin\Theta(\mU,\hat\mU).
\end{align}

\end{proof}

\section{Discussion and future work}\label{sec:con}
We have developed a supervised tensor decomposition method with side information on multiple modes. One important challenge of tensor data analysis is the complex interdependence among tensor entries and between multiple features. Our approach incorporates side information as feature matrices in the conditional mean tensor. The empirical results demonstrate the improved interpretability and accuracy over previous approaches. Applications to the brain connection and political relationship datasets yield conclusions with sensible interpretations, suggesting the practical utility of the proposed approach.
 
There are several possible extensions from the work. We have provided accuracy guarantees for parameter estimation in the supervised tensor model. Statistical inference based on tensor decomposition is an important future direction. Measures of uncertainty, such as confidence envelope for space estimation, would be useful. One possible approach would be performing parametric bootstrap~\citep{efron1994introduction} to assess the uncertainty in the estimation. 
For example, one can simulate tensors from the fitted low-rank model based on the estimates, and then assess the empirical distribution of the estimates. 
While being simple, bootstrap approach is often computationally expensive for large-scale data. Another possibility is to leverage recent development in debiased inference with distributional characterization~\citep{chen2019inference}. This approach has led to fruitful results for matrix data analysis. Uncertainly quantification involving tensors are generally harder, and establishing distribution theory for tensor estimation remains an open problem.

One assumption made by our method is that tensor entries are conditionally independent given the linear predictor $\Theta$. This assumption can be extended by introducing a more general mixed-effect tensor model. For example, in the special case of Gaussian model, we can model the first two moments of data tensor using
\begin{align}\label{eq:extention}
\mathbb{E}(\tY|\mX_1,\ldots,\mX_K)&=\tC\times \{\mM_1, \cdots ,\mM_K\},\\
 \text{Var}(\tY|\mX_1,\ldots,\mX_K)&=\mPhi_1\otimes \cdots \otimes \mPhi_K,
\end{align}
where $\mPhi_k\in\mathbb{R}^{d_k\times d_k}$ is the unknown covariance matrix on the mode $k\in[K]$. For general exponential family, an additional mean-variance relationship should also be considered. The joint estimation of mean model $\Theta$ and variance model $\mPhi_k$ will lead to more efficient estimation in the presence of unmeasured confounding effects. However, the introduction of unknown covariance matrices $\mPhi_k$ dramatically increases the number of parameters in the problem. { Suitable regularization such as graphical lasso or specially-structured covariance~\citep{li2017parsimonious,lock2018supervised} should be considered.} The extension of tensor modeling with heterogeneous mixed-effects will be an important future direction. 

Although we have presented the data applications in the context of order-3 data tensors, the framework of the supervised tensor decomposition applies to a variety of multi-way datasets. One possible application is the integrative analysis of omics data, in which multiple types of omics measurements (gene expression, DNA methylation, microRNA) are collected in the same set of individuals~\citep{lock2013joint,wang2019three}. Other applications include time-series tensor data with multiple side information. Exploiting the benefits and properties of  supervised tensor decomposition in specialized task will boost scientific discoveries.

\newpage
\section*{Appendix}

\appendix

\renewcommand{\thefigure}{S\arabic{figure}}
\setcounter{figure}{0}   
\renewcommand{\thetable}{S\arabic{table}}
\setcounter{table}{0}   

\section{Additional simulation results}\label{sec:asimulation}  

\subsection{Detailed simulation setup for Figure~\ref{fig:noniid}a-b}

We generate data from \textbf{Envelope} model \citep{li2017parsimonious} with slight modification. We simulate response tensor $\tY\in\mathbb{R}^{d\times d\times d}$ from the following model with envelope dimension $(u_1, u_2)$,
\begin{align}\label{eq:e}
    \tY|\mX &= \tB \times_3 \mX + \tE = \tC \times \{\mGamma_1, \mGamma_2,\mX\}+ \tE, \notag \\
    \text{with} \quad \tE &\sim \mathcal{TN}(\mSigma_1, \mSigma_2, \mI), \quad \mSigma_k = \mGamma_k \mOmega_k \mGamma_k^T + \mGamma_{0k} \mOmega_{0k} \mGamma_{0k}^T+\mI, \quad k = 1,2,
\end{align}
 where $\mX\in\mathbb{R}^{d\times p}$ is the feature matrix, $\tB=\tC\times\{\mGamma_1,\mGamma_2, \mI\}\in \mathbb{R}^{d\times d\times p}$ is the coefficient tensor, $\tC\in\mathbb{R}^{\mu_1\times \mu_2\times p}$ is a full-rank core tensor, $\mathcal{TN}(\cdot,\ \cdot,\ \cdot)$ represents zero-mean tensor normal distribution with Kronecker structured covariance, 
 $\mGamma_{k} \in \mathbb{O}(d,u_k)$ 
 consists of orthogonal columns, $\mGamma_{0k} \in \mathbb{O}(d,d-u_k)$ is the orthogonal complement of $\mGamma_k$, and  $\mOmega_k = \mA_k\mA_k^T$, $\mOmega_{0k} = \mA_{k0}\mA_{k0}^T$ with $\mA_k \in \bbR^{u_k \times u_k}$,  $\mA_{k0} \in \bbR^{(d - u_k) \times (d - u_k)}$.

The entries of $\mX$ are i.i.d.\ drawn from $\tN(0,1)$, the entries of $\mA_k$, $\mA_{k0}$ are i.i.d.\ drawn from $\textup{Uniform}[-\gamma, \gamma]$, and the entries of core tensor $\tC$ are i.i.d.\ drawn from $\textup{Uniform}[-3,  3]$. We call $\gamma$ the \textit{correlation level}. Note that the only distinction between model~\eqref{eq:e} and standard {\bf Envelope} model is the additional identity matrix $\mI$ in the expression of $\mSigma_k$. When $\gamma = 0$, the model~\eqref{eq:e} reduces to our \textbf{STD} model with rank $\mr = (u_1,u_2,p)$. We set $d=20, p=5$ in our simulation.  

\subsection{Detailed simulation setup for Figure~\ref{fig:noniid}c-d}

We generate the data from \textbf{GLSNet} model \citep{zhang2018network} with slight modification. We simulate the binary response tensor $\tY \in \{0,1\}^{d \times d \times d}$ from the following model
\begin{align}
    \mathbb{E}[\tY|\mX] = f({\bf 1} \otimes {\bf \Theta} + \tB \times_3 \mX),
\end{align}
where $f(\cdot)$ is the logistic link, $\mX \in \mathbb{O}(d,p)$ is the feature matrix with orthonormal columns, ${\bf \Theta}=\mA \mA^T \in \bbR^{ d \times d }$ is a rank-$R$  intercept matrix, where the entries of $\mA \in \bbR^{d \times R}$ are simulated from i.i.d.\ standard normal. 
Unlike original \textbf{GLSNet} model, we generate joint sparse and low-rank structure to the coefficient tensor $\tB$ as follows. 

To generate $\tB$, we firstly generate a low-rank tensor $\tB_0$ as 
\begin{align}
    \tB_0 = \tC \times \mM_1 \times \mM_2 \times \mM_3,
\end{align}
where $\tC \in \bbR^{R \times R \times R}$ is a full-rank core tensor, $\mM_1, \mM_2 \in \bbR^{d \times R}$ and $\mM_3 \in \bbR^{p \times R}$ are the factor matrices with orthonormal columns. We simulate i.i.d.\ uniform entries in $\tC$ and rescale the tensor $\tB_0$ such that $\onorm{\tB_0}_{\max} = 2$. Last, we obtain a sparse $\tB$ by randomly setting $sd^2p$ entries in $\tB_0$ to zero. We call $s$ the \textit{sparsity level} which quantifies the proportion of zero's in $\tB$. Hence, the generated tensor $\tB$ is of sparsity level $s$ and of low-rank $(R,R,R)$. We set $d = 20$, $p = 5$ and consider the combination of rank $R = 2$ (low), $4$ (high) and sparsity level $s = \{0, 0.3,0.5\}$ in the simulation.

\subsection{Comparison with GLMs under stochastic block models}
We investigate the performance of our model under correlated feature effects. We mimic the scenario of brain imaging analysis. A sample of $d_3=50$ networks are simulated, one for each individual. Each network measures the connections between $d_1=d_2=20$ brain nodes. We simulate $p=5$ features for the each of the 50 individuals. These features may represent, for example, age, gender, cognitive score, etc. Recent study has suggested that brain connectivity networks often exhibit community structure represented as a collection of subnetworks, and each subnetwork is comprised of a set of spatially distributed brain nodes. To accommodate this structure, we utilize the stochastic block model~\citep{abbe2017community} to generate the effect size. Specifically, we partition the nodes into $r$ blocks by assigning each node to a block with uniform probability. Edges within a same block are assumed to share the same feature effects, where the effects are i.i.d.\ drawn from $N(0,1)$. We then apply our tensor regression model to the network data using the BIC-selected rank. Note that in this case, the true model rank is unknown; the rank of a $r$-block network is not necessarily equal to matrix rank $r$~\citep{zeng2019multiway}. 

\begin{figure}[ht]
\centering
\includegraphics[width=15cm]{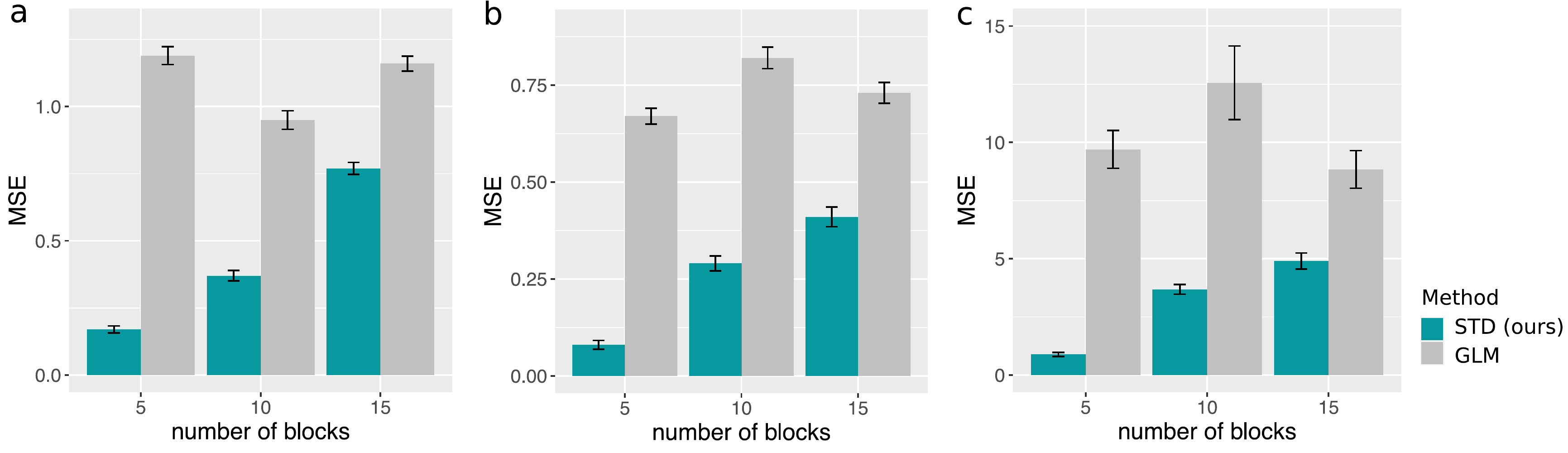}
\caption{Performance comparison under stochastic block models. The three panels plot the MSE when the response tensors are generated from (a) Gaussian (b) Poisson and (c) Bernoulli models. The $x$-axis represents the number of blocks in the networks. }\label{fig:glm}
\end{figure}

Figure~\ref{fig:glm} compares the MSE of our method with a multiple-response GLM approach. The multiple-response GLM is to regress the dyadic edges, one at a time, on the features, and this model is repeatedly fitted for each edge. As we find in Figure~\ref{fig:glm}, our tensor regression method achieves significant error reduction in all three data types considered. The outperformance is substantial in the presence of large communities; even in the less structured case ($\sim 20/15=1.33$ nodes per block), our method still outer-performs GLM. The possible reason is that the multiple-response GLM approach does not account for the correlation among the edges, and suffers from overfitting. In contrast, the low-rankness in our modeling incorporates the shared information across entries. By selecting the rank in a data-driven way, our method achieves accurate estimation in a wide range of settings.

\section{Additional results on data application}\label{sec:adata}
\subsection{Rank selection for Nations data}
Table~\ref{tab:rank_select} summarizes the BIC results in the grid search $\mr\in \{3,4,5\}^3$. We set $r_1=r_2$ due to the symmetry in the dataset. Table~\ref{tab:rank_select} shows that $(r_1,r_2)=(4,4)$ consistently provides the minimal BIC under a range of $r_3$. Because multiple values of $r_3$ give similar BIC, we choose $r_3$ based on the interpretability of the results. Tables~\ref{tab:s1_3}-\ref{tab:s1_5} compare the clustering results for $r_3=3,4,5$. For ease of visualisation, we list only the subset of relations for which the three configurations yield incoherent clustering. We find that the clustering with $r_3 = 4$ (Table~\ref{tab:s1_4}) provides the cleanest results. Table~\ref{tab:s1_3} with $r_3 = 3$ mixes the categories Economics with Organization and Military. Table~\ref{tab:s1_5} with $r_3 = 5$ mixes Economics with Organization, while splitting Military and Territory into different clusters. Therefore, we choose the rank $\mr=(4,4,4)$ in the main paper. The running time for the rank selection via grid search is 95 secs in total, on an iMac macOS High Sierra 10.13.6 with Intel Core i5 3.8 GHz CPU and 8 GB RAM. This indicates the BIC is feasible in the considered setting.

     \begingroup
     \setlength{\tabcolsep}{4.5pt} 
    \renewcommand{\arraystretch}{1} 
     \begin{table}[htb]
         \centering
         \begin{tabular}{c||ccc|ccc|ccc}
         $r_3$ & \multicolumn{3}{c|}{ $r_3 = 3$} & \multicolumn{3}{c|}{ $r_3 = 4$} & \multicolumn{3}{c}{ $r_3 = 5$}\\
         \hline
             $(r_1,r_2)$ & $(3,3)$ &  $(4,4)$ &  $(5,5)$ & $(3,3)$ &  $(4,4)$ &  $(5,5)$ &  $(3,3)$ &  $(4,4)$ &  $(5,5)$ \\
             \hline\hline
              BIC& 11364 & {\bf 11194} & 11701 & 12275 & {\bf 11897} & 12365 & 17652 & {\bf 12666} & 18146\\
         \end{tabular}
         \caption{BIC results for {\it Nations} data under different tensor rank. Bold number indicates the minimal BIC with a certain $r_3$. }
         \label{tab:rank_select}
     \end{table}
     \endgroup

     \begin{table}[h!]
    \resizebox{1\textwidth}{!}{%
    \begin{tabular}{c|l}
     Cluster& \multicolumn{1}{c}{Relations}\\
    \hline
    \multirow{2}{*}{ I}  & \textcolor{YellowOrange}{exportbooks}, \textcolor{YellowOrange}{relexportbooks}, \textcolor{OliveGreen}{protests}, \textcolor{YellowOrange}{tourism}, \textcolor{YellowOrange}{reltourism}, \textcolor{Blue}{relintergovorgs}\\
    & \textcolor{Blue}{relngo}, \textcolor{Blue}{intergovorgs3}, \textcolor{Blue}{ngoorgs3}, \textcolor{OliveGreen}{militaryalliance},\textcolor{OliveGreen}{commonbloc1} \\
    \hline
    \multirow{2}{*}{ II}  & \textcolor{OliveGreen}{militaryactions}, \textcolor{OliveGreen}{severdiplomatic},\textcolor{OliveGreen}{expeldiplomats}, \textcolor{OliveGreen}{commonbloc0},
    \textcolor{BrickRed}{aidenemy} \\ & \textcolor{BrickRed}{attackembassy},  \textcolor{BrickRed}{lostterritory},    \textcolor{OliveGreen}{blockpositionindex} \\
    \hline
    \multirow{2}{*}{III} & \textcolor{YellowOrange}{tourism3}, \textcolor{YellowOrange}{exports}, \textcolor{YellowOrange}{relexports}, \textcolor{YellowOrange}{exports3}, \textcolor{Blue}{intergovorgs}, \\ & 
    \textcolor{Blue}{ngo} ,\textcolor{OliveGreen}{ embassy}, \textcolor{OliveGreen}{reldiplomacy}, \textcolor{OliveGreen}{commonbloc2} \\
    \hline
     \multicolumn{2}{l}{ \footnotesize \fcolorbox{black}{YellowOrange}{\hspace{2mm}} 
     \raisebox{-1mm}{Economics} \quad   \fcolorbox{black}{OliveGreen}{\hspace{2mm}}  \raisebox{-1mm}{ Military}\quad \fcolorbox{black}{Blue}{\hspace{2mm}} \raisebox{-1mm}{Organization} \quad  \fcolorbox{black}{BrickRed}{\hspace{2mm}} \raisebox{-1mm}{Territory}}
    \end{tabular}
    }
    \caption{$K$-mean relations clustering with $r_3 = 3$. For visualization purpose, only a subset of relations are presented. See texts for details.}\label{tab:s1_3}
    \end{table}
    
    \begin{table}[H]
    \resizebox{1\textwidth}{!}{%
    \begin{tabular}{c|l}
     Cluster& \multicolumn{1}{c}{Relations}\\
    \hline
    \multirow{1}{*}{ I} &\textcolor{BrickRed}{aidenemy}, \textcolor{BrickRed}{attackembassy}, \textcolor{BrickRed}{lostterritory}\\
    \hline
    \multirow{2}{*}{ II}  &\textcolor{OliveGreen}{militaryactions}, \textcolor{OliveGreen}{severdiplomatic}, \textcolor{OliveGreen}{expeldiplomats}, \textcolor{OliveGreen}{protests}, \\
    & \textcolor{OliveGreen}{commonbloc0}, \textcolor{OliveGreen}{blockpositionindex}, \textcolor{OliveGreen}{commonbloc1}\\
    \hline
    \multirow{1}{*}{III} & \textcolor{Blue}{relintergovorgs}, \textcolor{Blue}{relngo}, \textcolor{Blue}{intergovorgs3}, \textcolor{Blue}{ngoorgs3}, \textcolor{OliveGreen}{militaryalliance}, \textcolor{OliveGreen}{commonbloc2}\\
    \hline
    \multirow{2}{*}{IV} &\textcolor{YellowOrange}{exportbooks}, \textcolor{YellowOrange}{relexportbooks}, \textcolor{YellowOrange}{tourism}, \textcolor{YellowOrange}{reltourism}, \textcolor{YellowOrange}{tourism3} \\
    &\textcolor{YellowOrange}{exports}, \textcolor{YellowOrange}{relexports}, \textcolor{YellowOrange}{exports3}, \textcolor{Blue}{intergovorgs}, \textcolor{Blue}{ngo}, \textcolor{OliveGreen}{embassy}, \textcolor{OliveGreen}{reldiplomacy}\\
    \hline
    \multicolumn{2}{l}{ \footnotesize   \fcolorbox{black}{YellowOrange}{\hspace{2mm}} \raisebox{-1mm}{Economics} \quad 
 \fcolorbox{black}{OliveGreen}{\hspace{2mm}}    \raisebox{-1mm}{Military}\quad \fcolorbox{black}{Blue}{\hspace{2mm}}  \raisebox{-1mm}{Organization} \quad  \fcolorbox{black}{BrickRed}{\hspace{2mm}}  \raisebox{-1mm}{Territory}}
    \end{tabular}
    }
    \caption{$K$-mean relations clustering with $r_3 = 4$. For visualization purpose, only a subset of relations are presented. See texts for details.}\label{tab:s1_4}
    \end{table}
    
    \begin{table}[H]
    \resizebox{\textwidth}{!}{%
    \begin{tabular}{c|l}
     Cluster& \multicolumn{1}{c}{Relations}\\
    \hline
    \multirow{2}{*}{ I} &\textcolor{YellowOrange}{exportbooks}, \textcolor{YellowOrange}{relexportbooks}, \textcolor{YellowOrange}{tourism}, \textcolor{YellowOrange}{reltourism}, \textcolor{YellowOrange}{tourism3}, \textcolor{YellowOrange}{exports}, \textcolor{YellowOrange}{relexports}, \textcolor{YellowOrange}{exports3}\\
    &\textcolor{Blue}{intergovorgs}, \textcolor{Blue}{relintergovorgs}, \textcolor{Blue}{ngo}, \textcolor{Blue}{relngo}, \textcolor{Blue}{intergovorgs3}, \textcolor{Blue}{ngoorgs3}, \textcolor{OliveGreen}{embassy}, \textcolor{OliveGreen}{reldiplomacy}\\
    \hline
    \multirow{1}{*}{ II}  &\textcolor{BrickRed}{attackembassy}\\
    \hline
    \multirow{1}{*}{III} & \textcolor{OliveGreen}{commonbloc0}, \textcolor{OliveGreen}{blockpositionindex}\\
    \hline
    \multirow{1}{*}{IV} & \textcolor{OliveGreen}{militaryalliance}, \textcolor{OliveGreen}{commonbloc2}\\
    \hline
    \multirow{2}{*}{V} & \textcolor{OliveGreen}{militaryactions}, \textcolor{OliveGreen}{severdiplomatic}, \textcolor{OliveGreen}{expeldiplomats}, \textcolor{BrickRed}{aidenemy}, \textcolor{BrickRed}{lostterritory},\\ & \textcolor{OliveGreen}{protests}, \textcolor{OliveGreen}{commonbloc1}\\
    \hline
    \multicolumn{2}{l}{ \footnotesize \fcolorbox{black}{YellowOrange}{\hspace{2mm}} \raisebox{-1mm}{Economics} \quad   \fcolorbox{black}{OliveGreen}{\hspace{2mm}}   \raisebox{-1mm}{Military}\quad \fcolorbox{black}{Blue}{\hspace{2mm}} \raisebox{-1mm}{Organization} \quad 
    \fcolorbox{black}{BrickRed}{\hspace{2mm}} \raisebox{-1mm}{Territory}}
    \end{tabular}
    }
    \caption{$K$-mean relations clustering with $r_3 = 5$. For visualization purpose, only a subset of relations are presented. See texts for details.}\label{tab:s1_5}
    \end{table}

\subsection{Comparison with unsupervised decomposition}\label{sec:compare1}

We compare the supervised vs.\ unsupervised decomposition in the {\it Nations} data analysis. Table~\ref{tab:unsup} shows the clustering results based on classical unsupervised Tucker decomposition without the feature matrices. Table~\ref{tab:s1} shows the clustering results based on supervised tensor decomposition ({\bf STD}). Compared with supervised decomposition, the unsupervised clustering loses some interpretation. Similar relations {\it exports} and {\it relexports}, {\it ngo} and {\it relngo} are separated into different clusters.

\begin{table}[!h]
    \resizebox{\textwidth}{!}{%
    \begin{tabular}{c|l}
     Cluster& \multicolumn{1}{c}{Relations}\\
    \hline
    \multirow{2}{*}{ I}  &\textcolor{YellowOrange}{economicaid}, \textcolor{YellowOrange}{releconomicaid}, \textcolor{YellowOrange}{exportbooks}, \textcolor{YellowOrange}{relexportbooks}, \textcolor{OliveGreen}{weightedunvote}, \textcolor{OliveGreen}{unweightedunvote}, \\
    &\textcolor{YellowOrange}{tourism}, \textcolor{YellowOrange}{reltourism}, \textcolor{YellowOrange}{tourism3}, \textcolor{YellowOrange}{exports}, \textcolor{Blue}{intergovorgs}, \textcolor{Blue}{ngo}, \textcolor{OliveGreen}{militaryalliance}\\
    \hline
    \multirow{2}{*}{ II}  & \textcolor{OliveGreen}{warning}, \textcolor{OliveGreen}{violentactions}, \textcolor{OliveGreen}{militaryactions}, \textcolor{OliveGreen}{duration}, \textcolor{OliveGreen}{severdiplomatic}, \textcolor{OliveGreen}{expeldiplomats}, \textcolor{YellowOrange}{boycottembargo}, \textcolor{BrickRed}{aidenemy}, \\
    &\textcolor{OliveGreen}{negativecomm}, \textcolor{Blue}{accusation}, \textcolor{OliveGreen}{protests}, \textcolor{OliveGreen}{unoffialacts}, \textcolor{BrickRed}{attackembassy}, \textcolor{Blue}{relemigrants}, \textcolor{OliveGreen}{timesincewar}, \textcolor{BrickRed}{lostterritory}, \textcolor{OliveGreen}{dependent}\\
    \hline
    III & \textcolor{OliveGreen}{timesinceally}, \textcolor{OliveGreen}{independence}, \textcolor{OliveGreen}{commonbloc0}, \textcolor{OliveGreen}{blockpositionindex}\\
    \hline
    \multirow{3}{*}{IV}  &\textcolor{YellowOrange}{treaties}, \textcolor{YellowOrange}{reltreaties}, \textcolor{YellowOrange}{officialvisits}, \textcolor{YellowOrange}{conferences}, \textcolor{YellowOrange}{booktranslations}, \textcolor{YellowOrange}{relbooktranslations}\\
    &\textcolor{OliveGreen}{negativebehavior}, \textcolor{Blue}{nonviolentbehavior}, \textcolor{Blue}{emigrants}, \textcolor{Blue}{emigrants3}, \textcolor{Blue}{students}, \textcolor{Blue}{relstudents}, \textcolor{YellowOrange}{relexports}, \textcolor{YellowOrange}{exports3}\\
    & \textcolor{Blue}{relintergovorgs}, \textcolor{Blue}{relngo}, \textcolor{Blue}{intergovorgs3}, \textcolor{Blue}{ngoorgs3}, \textcolor{OliveGreen}{embassy}, \textcolor{OliveGreen}{reldiplomacy}, \textcolor{OliveGreen}{commonbloc1}, \textcolor{OliveGreen}{commonbloc2}\\
    \hline
    \multicolumn{2}{l}{ \footnotesize \fcolorbox{black}{YellowOrange}{\hspace{2mm}} \raisebox{-1mm}{Economics} \quad   \fcolorbox{black}{OliveGreen}{\hspace{2mm}}   \raisebox{-1mm}{Military}\quad \fcolorbox{black}{Blue}{\hspace{2mm}} \raisebox{-1mm}{Organization} \quad 
    \fcolorbox{black}{BrickRed}{\hspace{2mm}} \raisebox{-1mm}{Territory}}
    \end{tabular}
    }
    \caption{Clustering of relations based on unsupervised tensor decomposition. }\label{tab:unsup}
    \end{table}

\begin{table}[!h]
\resizebox{\textwidth}{!}{%
\begin{tabular}{c|l}
 Category & \multicolumn{1}{c}{Relations}\\
\hline
\multirow{2}{*}{ I}   & \textcolor{OliveGreen}{warning}, \textcolor{OliveGreen}{violentactions}, \textcolor{OliveGreen}{militaryactions}, \textcolor{OliveGreen}{duration}, \textcolor{OliveGreen}{negativebehavior}, \textcolor{OliveGreen}{protests}, \textcolor{OliveGreen}{severdiplomatic} \\
&\textcolor{OliveGreen}{timesincewar}, \textcolor{OliveGreen}{commonbloc0}, \textcolor{OliveGreen}{commonbloc1}, \textcolor{OliveGreen}{blockpositionindex},  \textcolor{OliveGreen}{expeldiplomats}\\   
\hline
\multirow{2}{*}{II}& \textcolor{Blue}{emigrants}, \textcolor{Blue}{emigrants3}, \textcolor{Blue}{relemigrants}, \textcolor{Blue}{accusation}, \textcolor{Blue}{nonviolentbehavior}, \textcolor{Blue}{ngoorgs3}, \textcolor{OliveGreen}{commonbloc2}, \textcolor{Blue}{intergovorgs3}\\      
&\textcolor{YellowOrange}{releconomicaid}, \textcolor{Blue}{relintergovorgs}, \textcolor{Blue}{relngo},  \textcolor{Blue}{students}, \textcolor{Blue}{relstudents}, \textcolor{YellowOrange}{economicaid}, \textcolor{OliveGreen}{negativecomm}, \textcolor{OliveGreen}{militaryalliance} \\
\hline
\multirow{3}{*}{ III} & \textcolor{YellowOrange}{treaties}, \textcolor{YellowOrange}{reltreaties}, \textcolor{YellowOrange}{officialvisits}, \textcolor{YellowOrange}{exportbooks}, \textcolor{YellowOrange}{relexportbooks}, \textcolor{YellowOrange}{booktranslations}, \textcolor{YellowOrange}{relbooktranslations}\\
&\textcolor{YellowOrange}{boycottembargo}, \textcolor{OliveGreen}{weightedunvote}, \textcolor{OliveGreen}{unweightedunvote}, \textcolor{YellowOrange}{reltourism}, \textcolor{YellowOrange}{tourism}, \textcolor{YellowOrange}{tourism3}, \textcolor{YellowOrange}{exports}, \textcolor{YellowOrange}{exports3}  \\
&\textcolor{YellowOrange}{relexports}, \textcolor{Blue}{intergovorgs}, \textcolor{Blue}{ngo}, \textcolor{OliveGreen}{embassy}, \textcolor{OliveGreen}{reldiplomacy}, \textcolor{OliveGreen}{timesinceally}, \textcolor{OliveGreen}{independence}, \textcolor{YellowOrange}{conferences}, \textcolor{OliveGreen}{dependent}\\
\hline
IV &\textcolor{BrickRed}{aidenemy}, \textcolor{BrickRed}{lostterritory}, \textcolor{OliveGreen}{unoffialacts}, \textcolor{BrickRed}{attackembassy}\\
\hline
\multicolumn{2}{l}{ \footnotesize \fcolorbox{black}{YellowOrange}{\hspace{2mm}} \raisebox{-1mm}{Economics} \quad   \fcolorbox{black}{OliveGreen}{\hspace{2mm}}   \raisebox{-1mm}{Military}\quad \fcolorbox{black}{Blue}{\hspace{2mm}} \raisebox{-1mm}{Organization} \quad 
    \fcolorbox{black}{BrickRed}{\hspace{2mm}} \raisebox{-1mm}{Territory}}
\end{tabular}
}
\caption{Clustering of relations based on supervised tensor decomposition. }\label{tab:s1}
\end{table}

\subsection{How different are supervised vs.\ unsupervised factors in general?}\label{sec:compare2}
It is helpful to realize that the unsupervised and methods address different aspects of the problem. The unsupervised decomposition identifies factors that explain most variation in the tensor, whereas the supervised decomposition identifies factors that are most attributable to side features. 

We provide a simple example here for illustration.
\begin{example}
Consider the following data tensor $\tY$ and one-sided feature matrix $\mX$, 
\begin{align}
    \tY = \me_1\otimes \me_1 \otimes \me_1+10\me_2\otimes \me_2 \otimes \me_2, \quad \mX=\me_1,
\end{align}
where $\me_i=(0,\ldots,0, 1,0,\ldots,0)^T$ is the $i$th canonical basis vector in $\mathbb{R}^d$ for $i=1,2$. 
Now, consider the unsupervised vs.\ supervised decomposition of $\tY$ with rank $\mr = (1,1,1)$. 
Then, the top supervised and unsupervised factors are perpendicular to each other, 
\[
\mM_{\text{sup},k} \perp \mM_{\text{unsup},k}, \quad \text{for all } k=1,2,3,
\]
where $\mM_{\text{sup},k}$, $\mM_{\text{unsup},k}$ denote the mode-$k$ factors from supervised and unsupervised decompositions, respectively. 
\end{example}

\begin{rmk}
This example shows complementary information between factors from supervised vs.\ unsupervised decompositions. In general, one could construct examples such that these two methods return {\bf arbitrarily different} factors.   
\end{rmk}

\newpage
\setstretch{1.25}
\bibliographystyle{apalike}

\bibliography{tensor_wang}

\end{document}